# Donor-Acceptor Co-Adsorption Ratio Controls Structure and Electronic Properties of Two-Dimensional Alkali-Organic Networks on Ag(100)

B. Sohail[1], P.J. Blowey[2,3], L.A. Rochford[4], P.T.P. Ryan[3,5], D.A. Duncan[3], T.-L. Lee[3], P. Starrs [3,6], G. Costantini[2,4], D.P. Woodruff[2*], R.J. Maurer[1,2*]

*(1) Department of Chemistry, University of Warwick, Coventry CV4 7AL, UK*

*(2) Department of Physics, University of Warwick, Coventry CV4 7AL, UK*

*(3) Diamond Light Source, Harwell Science and Innovation Campus, Didcot, OX11 0DE, UK*

*(4) School of Chemistry, University of Birmingham, Birmingham B15 2TT, UK*

*(5) Department of Materials, Imperial College, London SW7 2AZ, UK*

*(6) School of Chemistry, University of St. Andrews, St. Andrews, KY16 9AJ, UK*

**Abstract**

The results are presented of a detailed combined experimental and theoretical investigation of the influence of coadsorbed electron-donating alkali atoms and the prototypical electron acceptor molecule TCNQ (7,7,8,8-tetracyanoquinodimethane) on the Ag(100) surface. Several coadsorption phases were characterised by scanning tunnelling microscopy, low energy electron diffraction, and soft-X-ray photoelectron spectroscopy. Quantitative structural data were obtained using normal incidence X-ray standing wave (NIXSW) measurements and compared with the results of density functional theory (DFT) calculations using several different methods of dispersion correction. Generally good agreement between theory and experiment was achieved for the quantitative structures, albeit with prediction of the alkali atom heights being challenging for some methods. The adsorption structures depend sensitively on the interplay of molecule-metal charge transfer and long-range dispersion forces, which are controlled by the composition ratio between alkali atoms and TCNQ. The large difference in atomic size between K and Cs has negligible effects on stability, whereas increasing the ratio of K:TCNQ from 1:4 to 1:1 leads to a weakening of molecule-metal interaction strength in favour of stronger ionic bonds within the two-dimensional alkali-organic network. A strong dependence of the work function on the alkali donor-TCNQ acceptor co-adsorption ratio is predicted.

* Email: r.maurer@warwick.ac.uk, d.p.woodruff@warwick.ac.uk

## 1. Introduction

The use of molecular films as components for organic electronics has attracted growing interest in recent years.[1] Thin films composed of organic molecules or polymers are already applied within commercially available technologies such as organic light emitting diodes (OLED)[2,3] and organic photovoltaics (OPVs)[4,5]**.** A significant challenge is posed by the interface formed between organic molecules and metal electrodes, the electronic properties of the interface having the potential to significantly affect device functionality. Charge injection barriers can, in particular, have a deleterious effect on device performance, so a crucial aspect of organic semiconductor design is to optimise the energy level alignment across the metal-organic interface. The energy level alignment is strongly influenced by the presence of interfacial dipoles that are modulated by many contributions such as charge transfer,[6] the 'push back effect',[7–10] bonding and intrinsic dipoles.[11]

A variety of model systems have been investigated to understand such effects at the atomic scale and the implications they have for the interface. These include, but are not limited to, the use of self-assembled monolayers (SAMs),[12–14] doping organic semiconductors with strong acceptor species,[15–18] or deposition of (sub)monolayers of strong acceptors (or donors) at metal surfaces.[19–24] These systems, in particular the last group, have provided insight into adsorption-induced substrate work function changes. However, it is still considered a challenge to improve reliably the energy level alignment across the interface[1,25–27]**.** Thus, to move away from a trial-and-error approach to optimise the fabrication of thin films and towards a more rational design process, we must establish a deeper understanding of the interactions at the organic-inorganic interface. While state-of-the-art electronic structure calculations have the ability to predict structure, stability and dipole formation at the interface, [28] their ability to do so accurately and robustly for a variety of systems needs to be carefully validated.

A potential route to tuning energy level-alignment at the interface is the coupling of strong, organic acceptor molecules co-adsorbed with alkali atoms to act as a spacer layer between the metal electrode and the organic semiconductor.[29–31] Coadsorption with alkali atoms will not only affect the interfacial dipole (and consequently the level alignment) but also the adsorption structure when forming 2D metal-organic frameworks (2D-MOF). Various phases of K atoms coadsorbed with perylenetetracarboxylic dianhydride (PTCDA) on Ag(111) have previously been studied.[32] The incorporation of K atoms into the overlayer leads to profound structural arrangements which were identified through STM, ST[H]M and DFT. Additionally, these phases were explored in a joint experimental and theoretical approach which comprehensively detailed the change in electronic properties as a function of K coverage.[33]

Understanding the parameters that lead to the formation of such phases, can aid the use of alkali doping techniques to control the surface properties of metal-organic interfaces. A particularly relevant molecule in this context is 7,7,8,8-tetracyanoquinodimethane (TCNQ), which is a prototypical electron acceptor that can be used as an additive to interfaces to lower the energy barrier for electron transfer from the metal into the semiconductor. TCNQ is also capable of forming highly conductive charge transfer salts in combination with suitable electron donor molecules.[34–38] Moreover, TCNQ adsorbed on Ag surfaces has been shown to form a diverse range of phases on the (111)[39,40] and (100)[41] terminations, some of which have been found to incorporate Ag metal adatoms into the adsorbate network, inducing the formation of a two-dimensional metal-organic framework (2D-MOF). In particular, when co-deposited with alkali metals, TCNQ forms a number of different two-dimensional networks as a function of overall coverage and alkali:TCNQ ratio, representing an ideal system to study the controllable synthetic parameters for 2D-MOF formation. Floris et al. [42], using non-dispersion-corrected density functional theory (DFT), first drew attention to the possibility that alkali atom coadsorption with TCNQ on Ag(100) could be used to tune the interfacial dipole and hence the barrier to charge transfer at a metal-organic interface.

The inherent versatility of TCNQ is exemplified on Ag(111) as it was shown that the network comprised of TCNQ and Ag adatoms formed by TCNQ adsorption,[39] is disrupted by deposition of K atoms and subsequent annealing.[43] This treatment results in the formation of a new phase with stoichiometry $K_2$TCNQ, which we have shown previously[43] to form a 2D organic salt with significant intra-layer cohesion and, compared to an adlayer network of TCNQ and Ag adatoms, a weakened interaction with the substrate. In addition, the presence of K also led to a significant modification of the work function and the electronic structure. It is, in particular, these modifications of the electronic structure that state-of-the-art density functional theoretical calculations seek to quantify[44].

Robust theoretical modelling is crucial to understand the stability and electronic properties of metal-organic interfaces. It is well established that long-range dispersion interactions must be accounted for to achieve accurate predictions of the structure and stability of such systems.[45–47] In cases of strong charge transfer, such as in these electron donor / acceptor pairs, long-range dispersion needs to account for the change in polarizability that occurs during charge transfer. We have previously studied this in the context of Ag(111)-$K_2$TCNQ and have found that the well-established surface screened Tkatchenko-Scheffler vdW method ($vdW^{surf}$) for DFT does not fully capture this effect. As a remedy to account for this charge transfer, we employed a rescaling scheme of the atomic polarizability and $C_6$ coefficient.[43,48] It has not yet been assessed if more recent beyond-pairwise dispersion correction methods, such as the non-local Many Body Dispersion method (MBD-NL),[49] are able to fully capture the interplay of charge transfer effects and long-range dispersion found for donor-acceptor networks at metal surfaces; this we address as part of this manuscript.

Here, we present the results of a detailed combined experimental and theoretical investigation of four different experimentally realised TCNQ/alkali co-adsorption phases on the Ag(100) surface. Two of these, with stoichiometry $K(TCNQ)_4$ and $Cs(TCNQ)_4$, are phases that were previously investigated in the purely computational study of Floris *et al.*[42]**,** while the existence of the $Cs(TCNQ)_4$ was previously identified by scanning tunnelling microscopy (STM).[50] Our experimental investigation includes not only the characterisation of several different co-adsorption phases on Ag(100) by STM, low energy electron diffraction (LEED) and soft X-ray photoelectron spectroscopy (SXPS), but also quantitative structural information on the heights of the coadsorbed species above the surface using normal incidence X-ray standing waves (NIXSW) measurements. The paper is structured as follows: we first describe the experimental fabrication and characterisation of the different 2D-MOF phases on Ag(100), before developing structural models that achieve quantitative agreement between NIXSW measurements and DFT predictions. Having established a full atomistic understanding of the 2D-MOF structures, we compare trends in the different 2D-MOF phases in terms of structure, stability, and electronic properties upon incorporation of different alkali atoms (K vs Cs) and upon changing the coadsorption ratio between donor and acceptor (1:4, 1:2, 1:1). We find that the coadsorption ratio sensitively affects the work function and the balance of interactions within the 2D-MOF and between substrate and 2D-MOF. This enables us to extract some general principles for the fabrication of 2D-MOF layers on metallic surfaces.

## 2. Methods

### Experimental Methods

Experimental characterisation of the co-adsorption phases of TCNQ with Cs and K (and also Na) on Ag(100) was performed using STM and low-current (microchannel plate) low energy electron diffraction (MCP-LEED) at room temperature in a UHV surface science chamber at the University of Warwick, but also by MCP-LEED and soft X-ray photoelectron spectroscopy (SXPS) in the UHV end-station of beamline I09 of the Diamond Light Source[51] used for the NIXSW measurements. Well-ordered clean Ag(100) samples were cleaned *in situ* by cycles of sputtering ($Ar^+$, 1keV) and annealing in both chambers. Single molecular monolayer structures were prepared by vacuum deposition of TCNQ from molecular beam epitaxy sources installed in the chambers, with alkali deposition being effected from resistively heated SAES 'getter' sources. STM images, recorded in constant current mode using electrochemically etched polycrystalline tungsten tips, were plane corrected and flattened using the open source image-processing software Gwyddion.[52] To form the alkali-TCNQ coadsorption phases, TCNQ was first deposited on a clean Ag(100) surface to form the commensurate $\begin{pmatrix} 1 & 4 \\ -3 & -1 \end{pmatrix}$ phase using the preparation method described previously.[41]

The alkali atoms were then deposited to increasing coverages onto the sample at room temperature. The highest K coverage ordered phase required annealing to ~300°C to produce a well-ordered surface (see Table S1).

For the NIXSW measurements (also at room temperature), the X-ray absorption probabilities of the C and N atoms of TCNQ, as well as at the co-adsorbed K and Cs atoms, were monitored by recording the intensity of the C 1s, N 1s, K 2p and Cs 3d photoelectron spectral peaks. These ‘hard’ X-ray spectra, as well as the high-resolution soft X-ray spectra, were collected using a VG Scienta EW4000 HAXPES hemispherical electron analyser mounted at 90° to the incident photon beam, while sweeping the photon energy through the (200) Bragg condition at near-normal incidence to the surface. Both the high-resolution SXP spectra (measured using ‘soft’ X-rays at photon energies of ~400-900 eV) and the ‘HAXPE’ spectra recorded in NIXSW experiments (photoemission spectra at the ‘harder’ X-ray energies corresponding to the Bragg (200) scattering condition at ~3040 eV) were fitted using the CasaXPS software package to allow chemical-state specific NIXSW data to be extracted. Fitting of the NIXSW absorption profiles to extract the structural parameters included taking account of the non-dipolar effects on the angular dependence of the photoemission, using values for the backward-forward asymmetry parameter *Q*[53], obtained from theoretical angular distribution parameters.[54]

## Computational Methods

The Fritz-Haber Institute *ab initio* molecular simulations package (FHI-aims)[55] was employed to perform density functional theory calculations. We use the generalised gradient approximation (GGA) exchange-correlation functional by Perdew, Burke and Ernzerhof (PBE)[56] coupled with dispersion correction schemes to account for van der Waals contributions to the total energy. We employed the surface-screened Tkatchenko-Scheffler van der Waals scheme (PBE+vdW$^{surf}$)[57] with and without the use of a manual rescaling procedure of the free atom polarizability and $C_6$ coefficient as reported by Blowey et al.[43] We also employed the recently proposed non-local many-body dispersion scheme (PBE+MBD-NL)[58] which requires no such rescaling.

The adsorption structures were modelled as a periodically repeated unit cell comprising a single unit mesh described by experimentally determined matrices of the substrate lattice vectors, the different unit mesh structures $\begin{pmatrix} 4 & 2 \\ -2 & 4 \end{pmatrix}$, $\begin{pmatrix} 4 & 3 \\ -3 & 4 \end{pmatrix}$, and $\begin{pmatrix} 6 & 3 \\ -3 & 6 \end{pmatrix}$ containing, respectively, one, two or four TCNQ molecule(s). The Ag(100) surface was modelled as a slab consisting of four atomic layers and separated from its periodic image by a vacuum gap exceeding 90 Å. The bottom two Ag layers were constrained to be fixed; all other atoms were free to move to optimise the structure.

The coordinates of the atoms in the bottom two layers of the Ag slab were constrained to the bulk truncated structure of Ag and the positions of the adsorbate and top two layers of the substrate were allowed to relax. In the case of PBE+vdW$^{surf}$, we exclude interactions between Ag atoms and define screened $C_6$ coefficients as provided by Ruiz et al.[59] The Brillouin zone was sampled with an 8×8×1 Monkhorst-Pack[60] k-grid and the geometries were optimised to below a force threshold of 0.025 eV Å$^{-1}$. FHI-aims employs an all-electron numeric atomic orbital basis with tiered default basis set and integration grid specification. All equilibrium structures were optimised with the default "light" basis set, followed by optimisation with "tight" basis. In the case of work function calculations, we made a small change to the truncation distance of the Ag atom (the "cut_pot" value) from default 4 Å to 6 Å. All output files have been deposited as a data set in the NOMAD repository 10.17172/NOMAD/2022.07.28-1 and are freely available.

## 3. Results and Discussion

### Experimental characterisation of co-adsorption phases

As previously reported in an STM investigation by Abdurakhmanova *et al.*[50], co-adsorption of Cs and TCNQ on Ag(100) leads to the formation of an ordered phase in which bright atomic-scale features attributed to Cs atoms are each surrounded by four TCNQ molecules arranged like the four vanes of a windmill. This type of 'windmill' ordering is seen in other TCNQ adsorption phases involving other coadsorbed metal atoms (e.g. Mn[61]) but also in the absence of co-deposited metal adatoms[20,62,63]. The ordering of this phase was reported to correspond to the matrix $\begin{pmatrix} 6 & 3 \\ -3 & 6 \end{pmatrix}$, while the STM images indicate a stoichiometry of $Cs(TCNQ)_4$. Our STM and LEED measurements confirm these observations, the constant tunnelling current STM image being shown in Fig. 1a while the experimental LEED pattern and its simulation are shown in Fig. S1. As shown in Fig. 1b, a $\begin{pmatrix} 6 & 3 \\ -3 & 6 \end{pmatrix}$ K/TCNQ coadsorption phase also exists and the STM images indicate identical lateral ordering of the TCNQ molecules to that in the comparable $Cs(TCNQ)_4$ phase. It is notable, however, that whereas the Cs atoms are imaged brightly in the STM (Fig. 1a), the K atoms are not imaged in this way. This inability to uniquely identify coadsorbed alkali metal adatoms from STM images has also been seen in alkali/TCNQ layers on Ag(111)[64,65] but we infer from the identical arrangement of the TCNQ molecules that this phase has the comparable $K(TCNQ)_4$ stoichiometry.

Two further K-TCNQ coadsorption phases were observed with increasing K coverage leading to indicated stoichiometries of $K(TCNQ)_2$ and KTCNQ, with associated matrices of $\begin{pmatrix} 4 & 3 \\ -3 & 4 \end{pmatrix}$ and $\begin{pmatrix} 4 & 2 \\ -2 & 4 \end{pmatrix}$, respectively; the corresponding STM images are shown in Figs. 1c and 1d respectively, while their LEED patterns are shown in Figure S1. Table S1 summarises the properties of all four detected coadsorption phases. Note that the ordered KTCNQ phase was only observed after annealing to ~300°C for a few minutes of a $K(TCNQ)_2$ surface onto which additional K had been deposited. It should be noted that the DFT study of Floris *et al.*[42] assumed that a $\begin{pmatrix} 6 & 3 \\ -3 & 6 \end{pmatrix}$ phase, shown experimentally to exist for Cs and TCNQ coadsorption on Ag(100)[50] also existed for coadsorption of Li, Na, K with TCNQ. Our results demonstrate that this assumption is valid for K but there appears to be no equivalent phase with coadsorbed Na, although a NaTCNQ $\begin{pmatrix} 4 & 2 \\ -2 & 4 \end{pmatrix}$ phase equivalent to the KTCNQ phase does exist[66]. No experiments were performed with Li-TCNQ coadsorption.

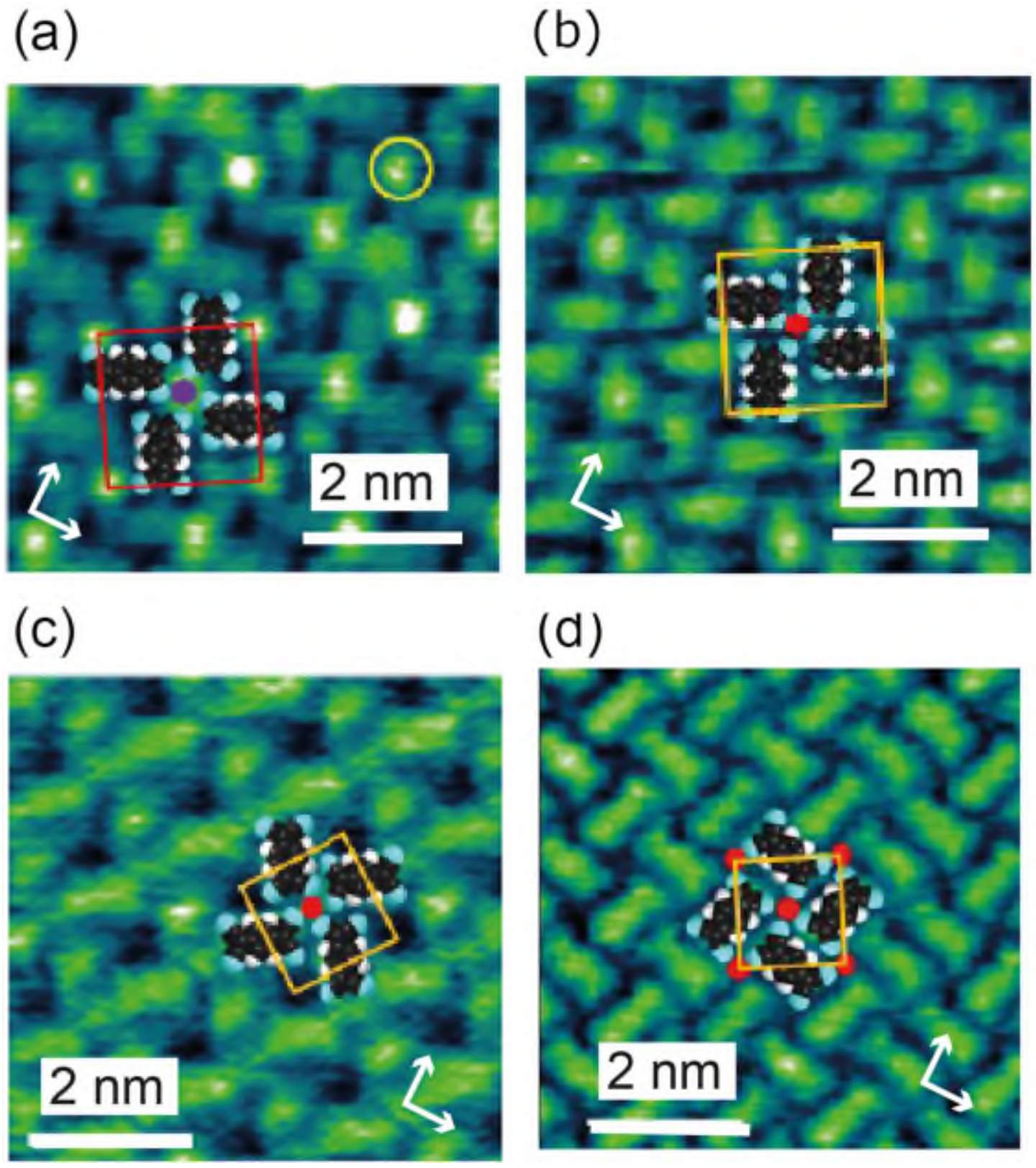

**Figure 1.** Experimental STM constant tunnelling current images of each of the investigated co-adsorbed phases.(a) $\begin{pmatrix} 6 & 3 \\ -3 & 6 \end{pmatrix}$ $Cs(TCNQ)_4$, (b) $\begin{pmatrix} 6 & 3 \\ -3 & 6 \end{pmatrix}$ $K(TCNQ)_4$, (c) $\begin{pmatrix} 4 & 3 \\ -3 & 4 \end{pmatrix}$ $K(TCNQ)_2$, (d) $\begin{pmatrix} 4 & 2 \\ -2 & 4 \end{pmatrix}$ KTCNQ. In each case the unit mesh and a simple schematic of the implied metal-organic structure are overlaid. Arrows show the <110> surface azimuthal directions while a bright feature attributed to a Cs atom is circled in (a). Tunnelling conditions of sample voltage and tunnelling current are (a) -1.3 V, 145 pA, (b) -0.9 V, 250 pA, (c) -1.0 V, 250 pA, (d) -0.6 V, 300 pA.

Additional characterisation of these coadsorption phases is provided by the SXP spectra, the spectra from the $K(TCNQ)_4$ phase being shown in Fig. 2. Similar spectra, recorded from the $Cs(TCNQ)_4$ and $K(TCNQ)_2$ phases, are presented in Fig. S2. No SXPS data were acquired from the KTCNQ phase. The absolute and relative photoelectron binding energies of the chemically distinct C 1s components are characteristic of a negatively charged TCNQ molecule[39,43] due to electron transfer from the substrate and/or the alkali atoms. The C 1s spectrum can be fitted with four unique features that are assigned to, in increasing binding energy, C atoms bound to H atoms (CH), aromatic C atoms bound only to other C atoms ($CC_1$), non-aromatic C atoms bound only to other C atoms ($CC_2$) and C atoms bound to N atoms (CN). Only a single N 1s peak was observed, indicating that the molecule remains intact with all N atoms in chemically similar bonding environments. Note that in the N 1s spectra two broad features are observed in the background; these relate to loss features from the neighbouring Ag 3d SXP peaks.

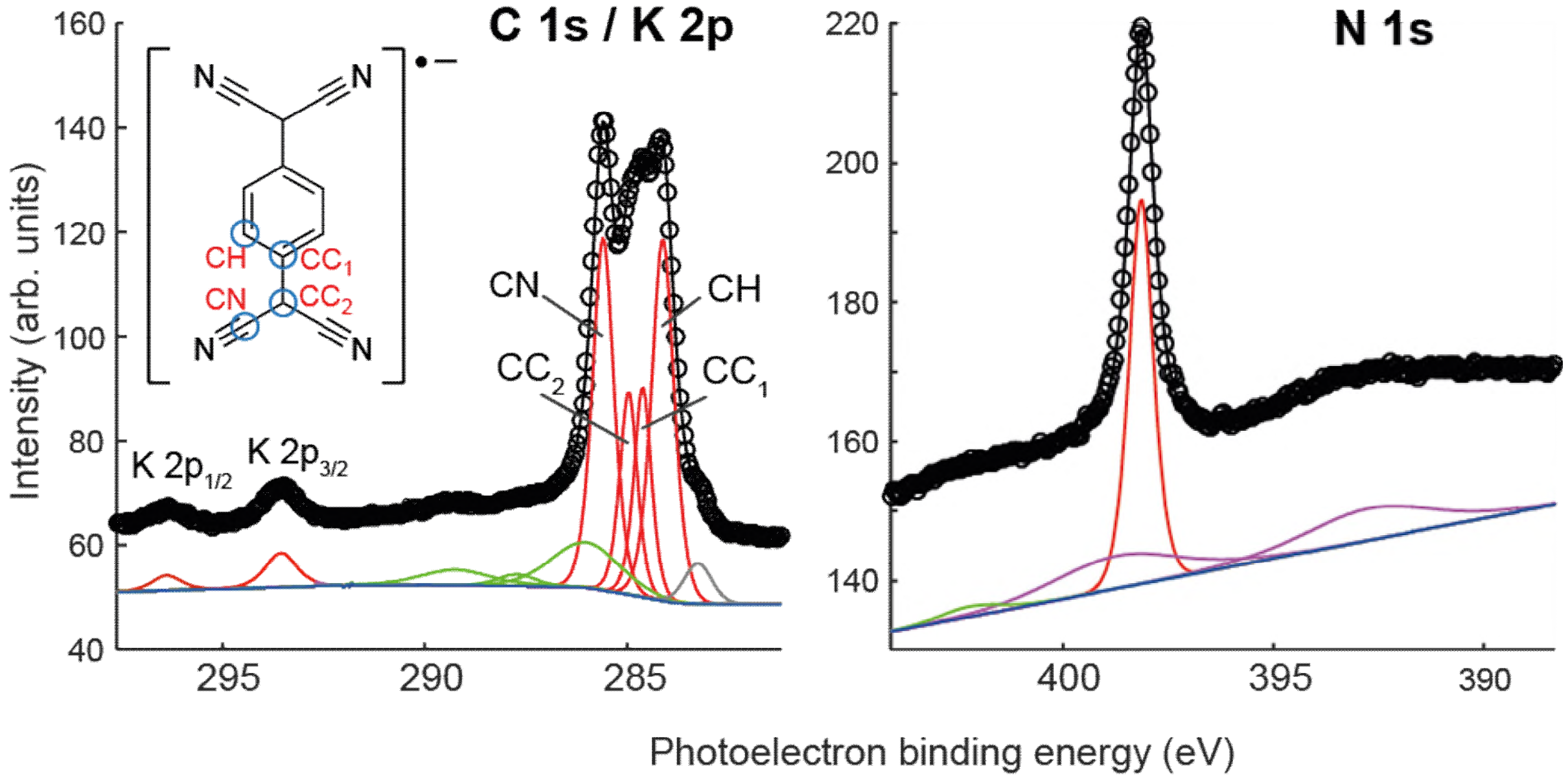

**Figure 2**. C 1s / K 2p and N 1s SXP spectra recorded from the $KTCNQ_4$ phase at photon energies of 435 eV and 550 eV, respectively. The main photoemission peaks (including the different chemically-shifted C 1s peaks) are shown in red. Satellites are shown in green while the loss satellites of the Ag 3d emission are shown in purple.

**Quantitative structure determination by NIXSW and DFT**

To complement this largely qualitative characterisation of these systems, we have undertaken experimental quantitative structural measurements using the normal-incidence X-ray standing wave technique (NIXSW)[53]. The general XSW technique[67] exploits the X-ray standing wave, created by the interference of an incident X-ray wave and the resulting Bragg-reflected wave, which shifts through the crystal as one scans across the Bragg condition by varying either the incidence angle or the photon energy. Monitoring the X-ray absorption rate profile of this standing wave in atoms of interest by measuring the element-specific core level photoemission allows one to determine the location of the absorbing atom relative to the Bragg diffraction planes.[68] If the Bragg diffraction planes coincide with the surface termination, as in the experiment presented here, the height of the absorbing atoms can be determined from the unique absorption profile that is acquired. Using core-level photoemission to monitor the absorption provides not only elemental specificity, but also chemical-state specificity, distinguishing the heights of C atoms in CH, CN and CC bonds in adsorbed TCNQ as well as determining the height of the Cs and K atoms. We have recently used this approach to investigate the nature of the $K_2$TCNQ co-adsorption phase on Ag(111), specifically demonstrating that this phase comprises a 2-D charge transfer salt.[43] The present study of phases of different alkali:TCNQ stoichiometry on Ag(100) broadens our insight into the nature of these co-adsorption phases.

The measured NIXSW absorption profiles from the *H* Bragg diffraction planes are fitted uniquely by two parameters, the coherent fraction, $f_H$, and the coherent position, $p_H$. In the idealised situation of absorber atoms in a single well-defined site with no static or dynamic disorder, $f_H$=1, and the coherent position corresponds to the absorber height above the extended Bragg diffraction planes in units of the Bragg plane spacing, $d_H$, leading to a height $D = (p_H + n) \cdot d_H$, where *n* is an integer (usually 0 or 1) chosen to ensure interatomic distances are physically reasonable. For Ag, $d_{(200)}$=2.04 Å. The influence of thermal vibrations of substrate and adsorbate atoms, and possible surface corrugation, can reduce the value of *f* by up to 20-30%, but values of *f* significantly lower than 0.70 must be attributed to two or more distinctly different contributing values of *D*.[69]

Table 1 shows the values of these structural parameters obtained from NIXSW measurements on the $K(TCNQ)_4$, $K(TCNQ)_2$, and $Cs(TCNQ)_4$ phases; no NIXSW measurements were acquired from the KTCNQ phase. Notice that at the higher photon energy of the (200) Bragg reflection (~3 keV compared to ~0.4 keV) the spectral resolution of the C 1s photoemission was not sufficient to resolve the distinct $CC_1$ and $CC_2$ components seen in the SXP spectrum of Fig. 2, so these were treated as a single component. All coherent fractions fall within the range consistent with the relevant atoms occupying a single well-defined height above the surface, with the notable exception of the CN and N atoms in the $K(TCNQ)_2$ phase. The low *f* value for the N atoms in this case clearly indicates that these N atoms occupy at least two distinctly different heights above the surface, in which case the corresponding *D* value is a weighted mean of the true heights. A height variation of the N atoms must lead to the same qualitative effect, but weaker, for the CN atoms bonded to the N atoms. The heights of all the molecular components in the $KTCNQ_4$ phase are essentially identical to those in $Cs(TCNQ)_4$, (and to the values for pure TCNQ adsorption[37]) although in the $K(TCNQ)_2$ phase the central quinoid ring appears to be about 0.1 Å higher.

**Table 1.** Experimental NIXSW (200) structural parameter values, the coherent fraction *f* and the coherent position, here converted to a height value *D,* for the three coadsorption phases investigated by this technique. Precision estimates (in units of the least significant figure) are shown in parentheses.

| **$K(TCNQ)_4$** | | | **$K(TCNQ)_2$** | | | **$Cs(TCNQ)_4$** | | |
|---|---|---|---|---|---|---|---|---|
| | *f* | *D* (Å) | | *f* | *D* (Å) | | ***f*** | *D* (Å) |
| CH | 0.68(10) | 2.72(5) | CH | 0.70(10) | 2.81(5) | CH | 0.63 (10) | 2.74(5) |
| CC | 0.83(10) | 2.64(5) | CC | 0.70(10) | 2.66(5) | CC | 0.80(10) | 2.63(5) |
| CN | 0.69(10) | 2.53(5) | CN | 0.53(10) | 2.69(5) | CN | 0.71(10) | 2.53(5) |
| N | 0.76(10) | 2.38(5) | N | 0.32(10) | 2.57(5) | N | 0.78(10) | 2.38(5) |
| K | 0.76(10) | 3.75(10) | K | 0.76(10) | 3.61(5) | Cs | 0.74(10) | 4.08(5) |

Notice, that the experimental precision estimates are random errors arising from the computational fitting and take no account of possible systematic errors. An additional systematic error may be associated with the *D* value for the K atoms in the $K(TCNQ)_4$ phase due to the difficulty of separating the weak K 2p emission from the underlying C 1s satellite features. The coverage of K atoms in this adsorption phase is extremely low (0.028 ML where 1 ML corresponds to one atom or molecule per surface layer Ag atom). Of course, the coverage of Cs atoms in the $Cs(TCNQ)_4$ phase has this same low value, and the Cs 3d peaks overlap the (broad) plasmon loss feature of the Ag 3s emission, but extrapolation of tabulated photoionization cross-sections[70] indicates that the value for the Cs 3d state is an order of magnitude larger than that of the K 2p state at the (200) Bragg reflection energy of ~3 keV; separation of the Cs 3d and the Ag 3s loss peaks is also eased by the fact that the Cs 3d peaks are much narrower.

Long-range dispersion-corrected DFT calculations were performed for all four of the coadsorption phases of Fig. 1 to establish the optimum structural parameters of the models, which can be compared to our experimental NIXSW measurements. These DFT calculations employed the PBE functional[71] to evaluate exchange-correlation accompanied by three flavours of dispersion schemes: the standard vdW$^{surf}$ method[59], the vdW$^{surf}$ method in which the $C_6$ coefficients of Cs and K have been rescaled to account for the change in atomic polarizability upon cation formation (vdW$^{surf}$($Cs/K^+$)), and the non-local many body dispersion method[49] (PBE+MBD-NL) as implemented in FHI-aims.[72] Complete comparisons of the experimental and predicted NIXSW parameters for each adsorption structure can be found in Tables S2-S5 of the supporting information. The optimised adsorption structures determined by the DFT calculations for the coadsorption systems on Ag(100) investigated here (shown in Figure 3) involve an alkali atom (Cs or K) coadsorbed with different numbers of TCNQ molecules, according to the experimentally inferred stoichiometries of 1:4 in ($Cs(TCNQ)_4$ and $K(TCNQ)_4$), 1:2 in ($K(TCNQ)_2$) and 1:1 in KTCNQ, respectively. The adsorption motif of the TCNQ molecules is reminiscent of windmill vanes lying on the substrate with the alkali atom at the centre, four-fold co-ordinated by cyano groups of neighbouring TCNQ molecules.

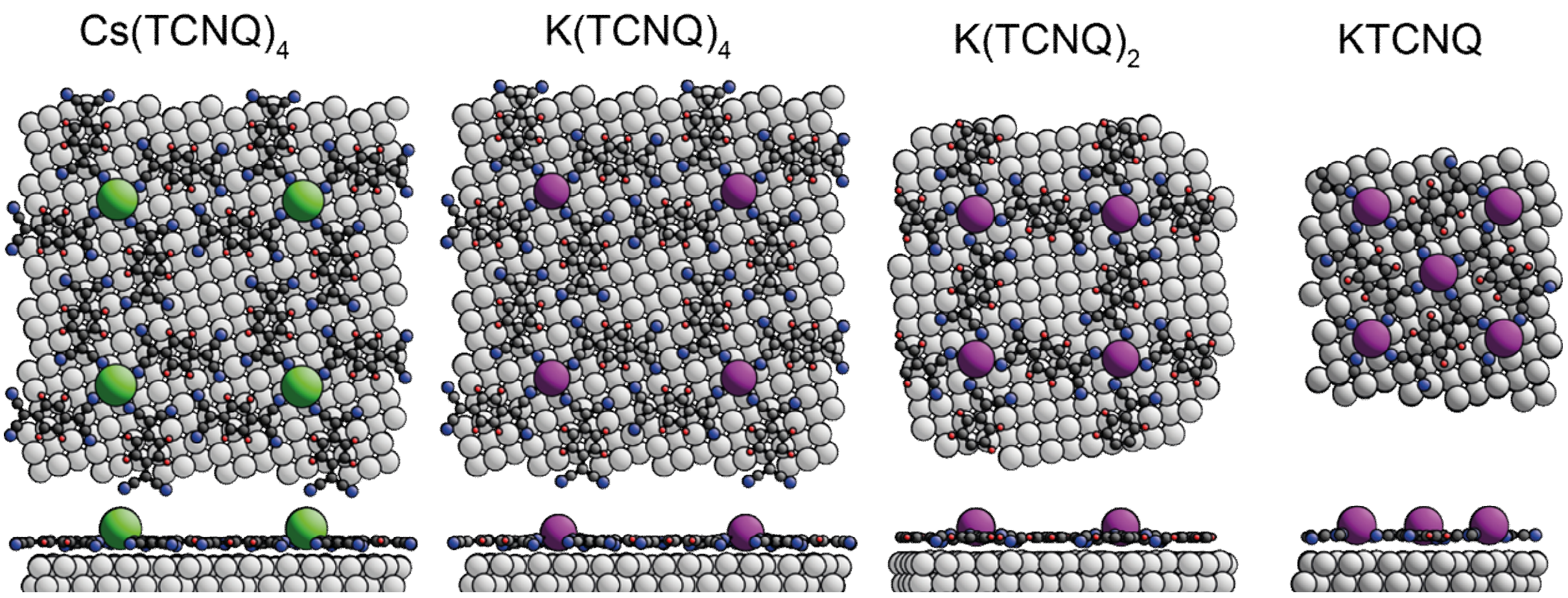

**Figure 3.** Plan and side view of the four coadsorbed phases as predicted by PBE+vdW$^{surf}$($Cs/K^+$): $Cs(TCNQ)_4$, $K(TCNQ)_4$, $K(TCNQ)_2$, and KTCNQ on Ag(100). Cs atoms are indicated by the green spheres. K atoms are indicated by purple spheres.

In the case of $Cs(TCNQ)_4$ and $K(TCNQ)_4$, our findings are in accordance with the structures proposed by Floris and colleagues[42] for $M(TCNQ)_4$ (M=Li,Na,K,Cs). Three of the peripheral cyano groups of each molecule are bent down toward the surface with the cyano group closest to the alkali atom pointing slightly upward towards the alkali atom, leading to N atoms having two different heights above the surface. The same structural feature is found in the $K(TCNQ)_2$ phase, but in this phase, there is an equal number of 'up' and 'down' N atoms, leading to a more significant decrease in the N coherent fraction value. Alkali atoms are four-fold coordinated to cyano groups in all the 1:4 and 1:2 stoichiometry phases, the main difference being that TCNQ molecules bridge neighbouring alkali atoms in the $K(TCNQ)_2$ phase, the alkali atom contacting two cyano groups on opposite ends of the molecule. The KTCNQ phase, in which K and TCNQ are adsorbed in a 1:1 ratio, is more closely packed and rigid relative to other phases as all four cyano groups of each molecule are in direct contact with a K atom.

**Table 2.** Comparison of the experimentally determined values of the NIXSW coherent position, converted into atomic heights, with the atomic heights relative to the average outermost Ag(100) layer found in the results of DFT calculations for the three alkali/TCNQ adsorption phases investigated. Precision estimates (in units of the least significant figure) are shown in parentheses. All theoretical values reported are calculated at the PBE+vdW$^{surf}$ (Cs/K$^+$) and PBE+MBD-NL level. No experimental NIXSW data exists for the KTCNQ/Ag(100) phase.

| | $Cs(TCNQ)_4$ | | | $K(TCNQ)_4$ | | |
|---|---|---|---|---|---|---|
| Species | *D* (Å) – Exp | *D* (Å) – PBE+ vdW$^{surf}$ (Cs/K+) | *D* (Å) – PBE+ MBD-NL | *D* (Å) – Exp | *D* (Å) – PBE+ vdW$^{surf}$ (Cs/K+) | *D* (Å) – PBE+ MBD-NL |
| C-H | 2.74(5) | 2.69 | 2.75 | 2.72(5) | 2.69 | 2.74 |
| C-C | 2.63(5) | 2.64 | 2.69 | 2.64(5) | 2.63 | 2.68 |
| C-N | 2.53(5) | 2.55 | 2.60 | 2.53(5) | 2.56 | 2.60 |
| N | 2.38(5) | 2.45 | 2.46 | 2.38(5) | 2.47 | 2.46 |
| K/Cs | 4.08(5) | 4.06 | 3.76 | 3.75(10) | 3.50 | 3.40 |
| | $K(TCNQ)_2$ | | | KTCNQ | | |
| C-H | 2.81(5) | 2.68 | 2.71 | - | 2.78 | 2.84 |
| C-C | 2.66(5) | 2.64 | 2.67 | - | 2.79 | 2.88 |
| C-N | 2.69(5) | 2.63 | 2.66 | - | 2.77 | 2.92 |
| N | 2.57(5) | 2.59 | 2.64 | - | 2.74 | 2.20 |
| K | 3.61(5) | 3.71 | 3.54 | - | 3.75 | 3.64 |

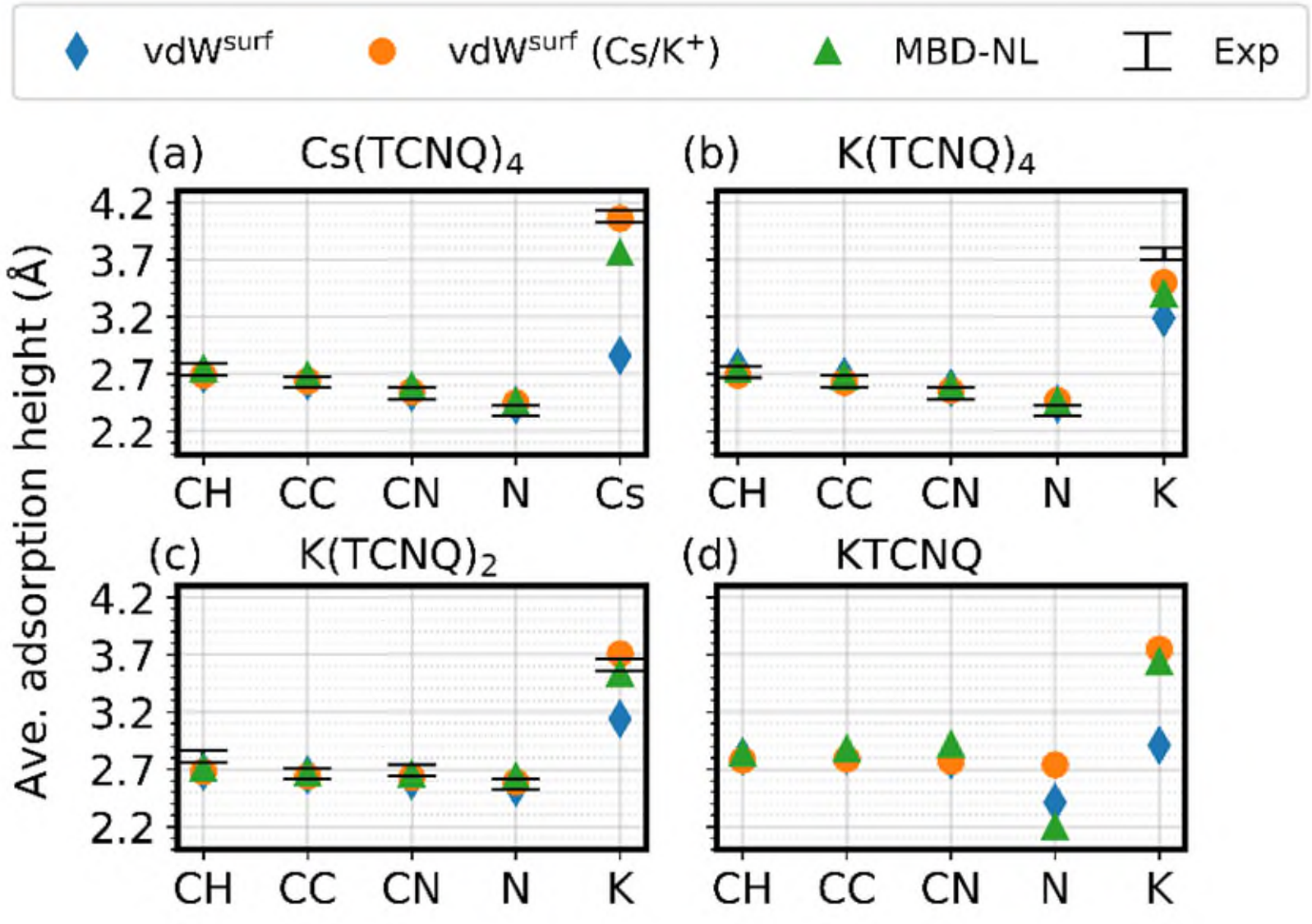

**Figure 4.** Atomic heights derived from dispersion inclusive DFT calculations using the three separate dispersion schemes employed in this work. The experimental NIXSW values are shown with black bars indicating the estimated error ranges.

Summarised in Table 2 and depicted in Figure 4 is a comparison of the coherent positions (expressed as heights) for the alkali atoms, the N atoms and the experimentally distinguishable carbon atoms within the TCNQ molecules (defined in Figure 2) determined experimentally by NIXSW and calculated by DFT using different long-range dispersion schemes. Agreement between the experimental and computed *D* values within the experimentally estimated precision is generally good. A comparison of the measured and calculated coherent fractions is shown in Tables S2 to S5. Notice that the $f_{theory}$ values shown in these tables are determined only by the small variations in height of symmetrically inequivalent atoms in the optimised structures; they take no account of static and dynamic disorder, so the measured values can be expected to be up to ~30% smaller than the theoretical values due to the effects of static and dynamic disorder perpendicular to the surface.[69] While most of the experimental and predicted values of the coherent fractions are consistent with a single height, or only a narrow range of heights, of the contributing atoms, in the $K(TCNQ)_2$ phase, the significantly lower measured coherent fraction value for the N atoms (0.32±0.10) suggests that there must be significant occupation of at least two different heights of these atoms. If one assumes that the coherent fraction for the N atoms includes a decrease of up to 30% due to static and dynamic disorder, then the maximum expected coherent fraction that could be predicted by the DFT simulations would be 0.46±0.14. With this in mind, the DFT simulations reproduce this expected coherent fraction well (0.58 and 0.45 with vdw$^{surf}$(K$^+$) and MBD-NL, respectively). Indeed, the DFT structural models for this phase do show equal occupation of two contributing heights of the N atoms differing by 0.52-0.72 Å in the different methods of dispersion correction. This height variation arises as N atoms that are adjacent to, and bonding to, the alkali atoms sit higher above the surface than N atoms that are not adjacent to alkali atoms. The range of different N atom heights predicted for the different phases by the two alternative DFT methods are shown in Figure 5. The MBD-NL method always predicts a broader range of N atom heights than the vdW$^{surf}$ method, despite both methods predicting very similar average adsorption heights.

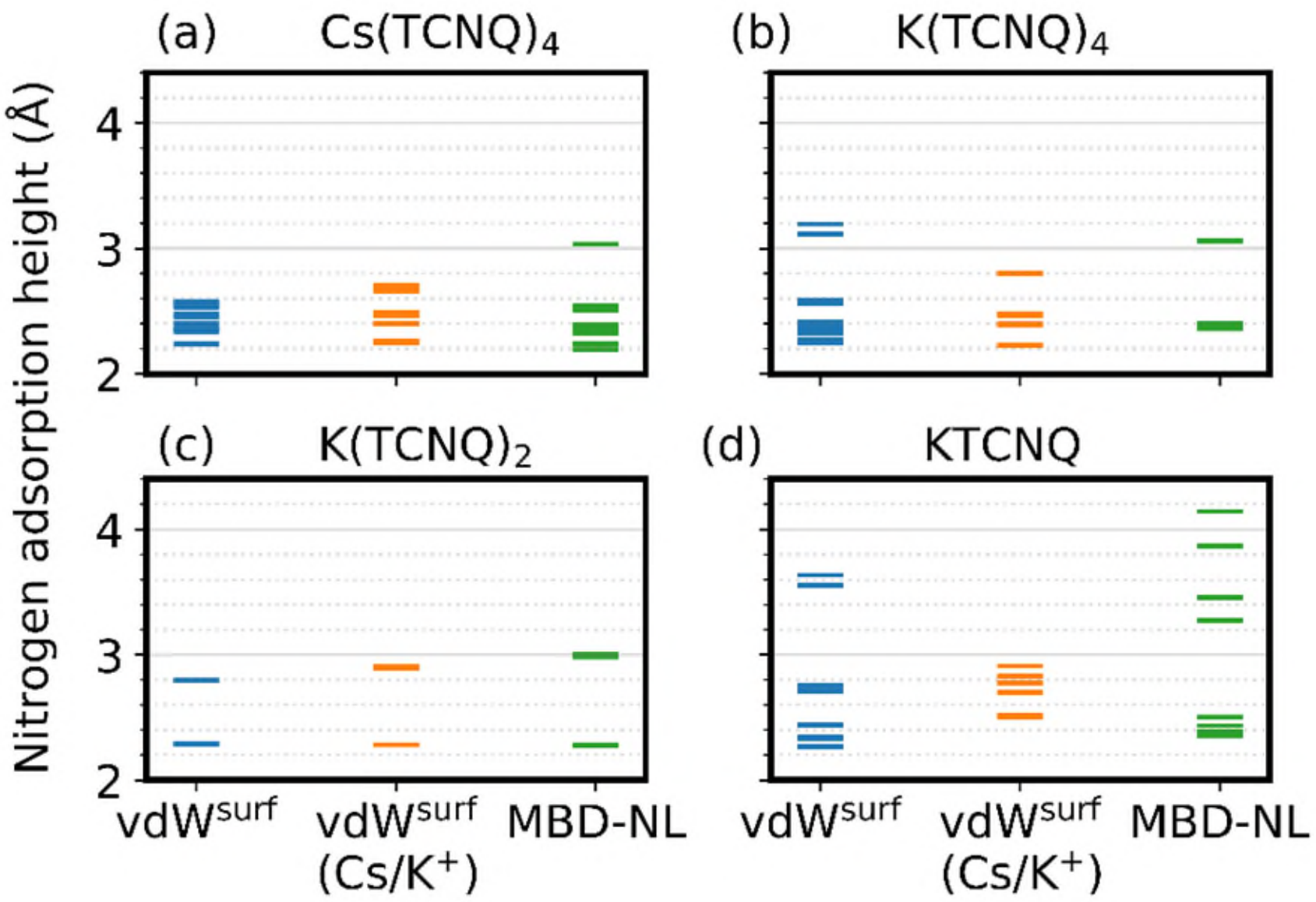

**Figure 5** Distribution of nitrogen atom adsorption heights as predicted by different dispersion-corrected DFT methods for (a) $Cs(TCNQ)_4$, (b) $K(TCNQ)_4$, (c) $K(TCNQ)_2$, and (d) KTCNQ. Each bar corresponds to the adsorption height of one nitrogen atom.

All the dispersion correction schemes show good agreement with the experimental NISXW layer spacings for the C and N species, as shown in Table 2 and Tables S2-S5 that present theoretical heights obtained from all three methods. However, the alkali atom adsorption height proved to be much more sensitive to the dispersion scheme employed. Figure 4 displays the height of different atomic species for all the coadsorption phases investigated here, derived from theoretical calculations employing the three different dispersion schemes together with the experimental values. The values derived from PBE+vdW$^{surf}$ show the worst agreement in alkali atom adsorption height compared to the experimental values. In all cases, the rescaled scheme PBE+vdW$^{surf}$(Cs/K$^{+}$) shows the best agreement for all the chemically distinct species. Notably, the Cs height is in good agreement with the experimental value due to our procedure of rescaling the relevant $C_6$ dispersion coefficient for the alkali atom; in the absence of rescaling (i.e. for the standard PBE+vdW$^{surf}$ scheme) the Cs height is 1.22 Å lower than experiment. Similarly, the K height in phases K(TCNQ)$_4$ and K(TCNQ)$_2$ without rescaling are 0.56 Å and 0.79 Å lower than experiment; these discrepancies are significantly reduced when rescaling is employed (to 0.25 Å and 0.10 Å, respectively). PBE+MBD-NL outperforms the default version of vdW$^{surf}$, but not the vdW$^{surf}$(Cs/K$^{+}$) approach. It determines the structural parameters in good agreement with our rescaled and experimental results for the molecular constituent atoms, but also for alkali atom heights. However, an important feature of PBE+MBD-NL is that it includes beyond-pairwise dispersion interactions, which are able to correct for the known overestimation of adsorption energies of PBE+vdW$^{surf}$.[45,73] In the case of PBE+vdW$^{surf}$(Cs/K$^{+}$), the manual parameter rescaling accounts for the effect of charge transfer on atomic polarizability. The MBD-NL method supposedly automatically accounts for this effect.

**Effect of overlayer composition on stability**

In our earlier study of co-adsorption of K atoms with TCNQ on the Ag(111) substrate[43], we found that the overlayer takes the form of a two-dimensional salt rather than a strongly surface-bound adlayer. This conclusion was reached through a decomposition of the energetic driving forces of adlayer formation and an analysis of the charge distribution. Here, we explore the extent to which this same effect is present on Ag(100) in the four different studied coadsorption phases. Specifically, we ask: what is the effect of changing the coadsorbed alkali atom (Cs vs. K), and what is the effect of changing the alkali:TCNQ ratio? These provide 'experimentally tuneable parameters', so it is important to know how they affect the stability and electronic properties of the interface.

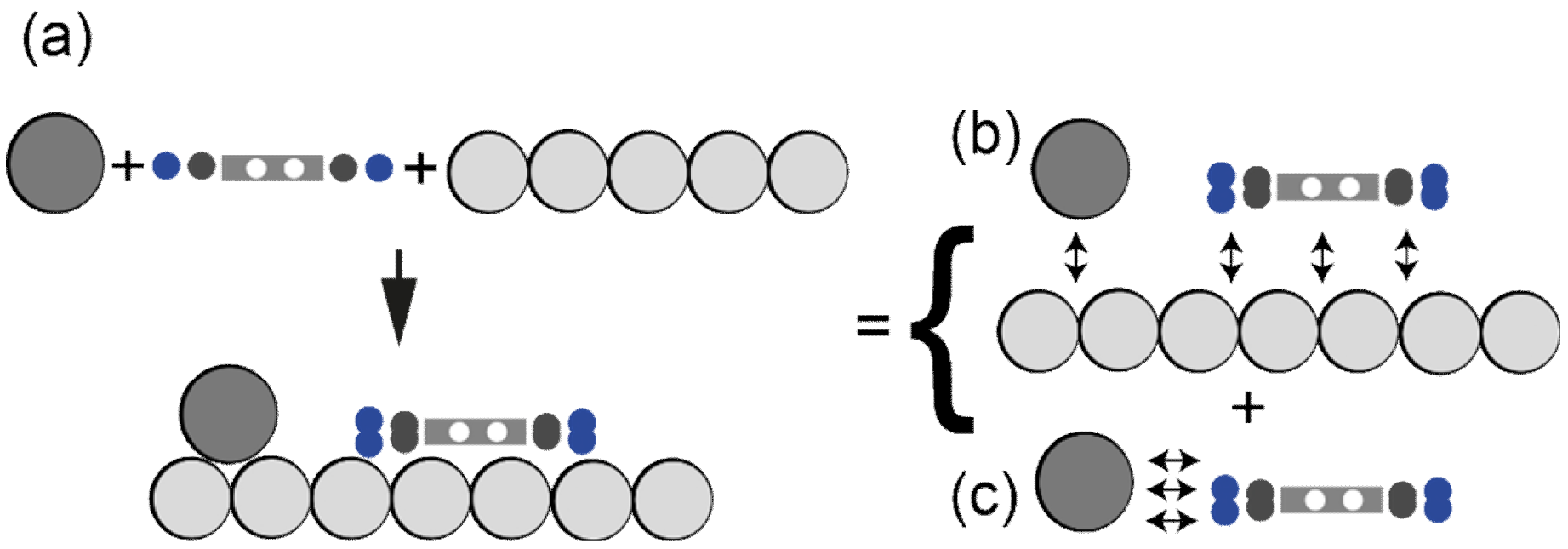

**Figure 6** Schematic depiction of the energetic contributions to the adsorption energy of the different adsorbates on Ag(100). Panel (a) shows the 3 components of the total co-adsorbed system: molecule, alkali atom (dark grey) and metal substrate, before and after co-adsorption (below arrow). Panel (b) shows a schematic representation of the adsorbate overlayer to substrate interaction, while (c) depicts the intra-adsorbate interaction. The double headed arrows indicate the interactions that are included in each contribution. The adsorbed molecules are shown with both ends twisted, as is found to be the case in $K(TCNQ)_2$.

Figure 6(a) shows a schematic representation of the components of each coadsorption phase on Ag(100), namely alkali atoms, molecular adsorbates, and the silver surface. Within the network we can decompose the total adsorption energy into two contributions: the adsorbate to substrate interaction and the intra-adsorbate interactions, shown in Figure 6(b) and (c), respectively. The total adsorption energy, $\mathrm{E_{ads}}$, was calculated using equation 1 (below), in which $\mathrm{E_{tot}}$, is the total energy of the co-adsorbed network, $\mathrm{E_{adlayer}}$ is the total energy of the combined alkali and TCNQ adsorbate layer (frozen in the adsorption geometry as a freestanding layer with the surface removed), $\mathrm{E_{surface}}$ is the total energy of the optimised bare Ag(100) surface, $\mathrm{E_{Cs,K}}$ is the total energy of the isolated neutral gas-phase alkali atom and $\mathrm{E_{TCNQ}}$ is the total energy of a relaxed gas-phase TCNQ molecule. The first term in square brackets in Equation 1 is the strength of adlayer-substrate interaction (as shown in Figure 6(b)), while the second term in square brackets describes the cohesive energy of the free-standing alkali-molecule layer (shown in Figure 6(c)). Both terms together yield the adsorption energy associated with bringing all components together.

$$\mathrm{E_{ads}} = \left[ (\mathrm{E_{adlayer}} + \mathrm{E_{surface}}) - \mathrm{E_{tot}} \right] + \left[ (\mathrm{E_{Cs,K}} + \mathrm{E_{TCNQ}}) - \mathrm{E_{adlayer}} \right] \quad (1)$$

**Table 3.** Total adsorption energies, $E_{ads}$, per unit surface area for each system calculated at PBE+vdW$^{surf}$(Cs/K$^+$) and PBE+MBD-NL levels.

| | $Cs(TCNQ)_4$ | $K(TCNQ)_4$ | $K(TCNQ)_2$ | KTCNQ |
|---|---|---|---|---|
| PBE+vdW$^{surf}$ (Cs/K$^+$) (eV nm$^{-2}$) | 4.95 | 4.98 | 5.39 | 8.57 |
| PBE+MBD-NL (eV nm$^{-2}$) | 4.16 | 4.17 | 4.60 | 7.61 |

The results, summarised in Table 3, show the total adsorption energies per unit surface area of the co-adsorbed networks calculated at both the PBE+vdW$^{surf}$(Cs/K$^+$) and PBE+MBD-NL level of theory. While both methods predict the same trends, the MBD-NL adsorption energies are considerably reduced compared to the pairwise scheme. Irrespective of the level of theory, the $K(TCNQ)_4$ and $Cs(TCNQ)_4$ phases are basically isoenergetic within our numerical tolerance. By increasing the alkali concentration from 1:4, to 1:2, to 1:1, the adsorption energies increase monotonically, with the 1:1 phase being the most energetically stable.

Figure 7 shows a histogram of the absolute adsorption energies per unit surface area broken down into the two contributions shown in equation 1. Both 1:4 networks exhibit the same breakdown with the adsorbate-substrate interaction presenting the largest contribution to the overall adsorption energy (around 70%). This is predicted by both PBE+vdW$^{surf}$(Cs/K$^+$) (Figure 7(a)) and PBE+MBD-NL (Figure 7(b)) methods. This can be attributed to the fact that $Cs(TCNQ)_4$ and $K(TCNQ)_4$ show similar optimised adsorption structures (see Figure 2 and Table 1), with similar lateral and perpendicular arrangements at the surface. Both networks adopt a TCNQ windmill vane motif surrounding a central alkali atom. In addition, the alkali atom in both phases occupies the same hollow site with respect to the underlying surface. (Alternative registries were explored in the DFT calculations but were energetically less favourable and failed to account for the discrepancy between the calculated and experimentally determined K atom height in the $K(TCNQ)_4$ phase). In the lower panels of Figure 7, the average adsorption height of TCNQ molecules within the networks is shown as a function of the alkali:TCNQ stoichiometry and choice of alkali atom. In both $M(TCNQ)_4$ phases and the $K(TCNQ)_2$ phase, the cyano groups are twisted. The N atoms close to an alkali atom are higher above the surface than those not adjacent to an alkali atom, which are bent downwards towards the underlying Ag(100) surface to maximise interactions with the substrate.

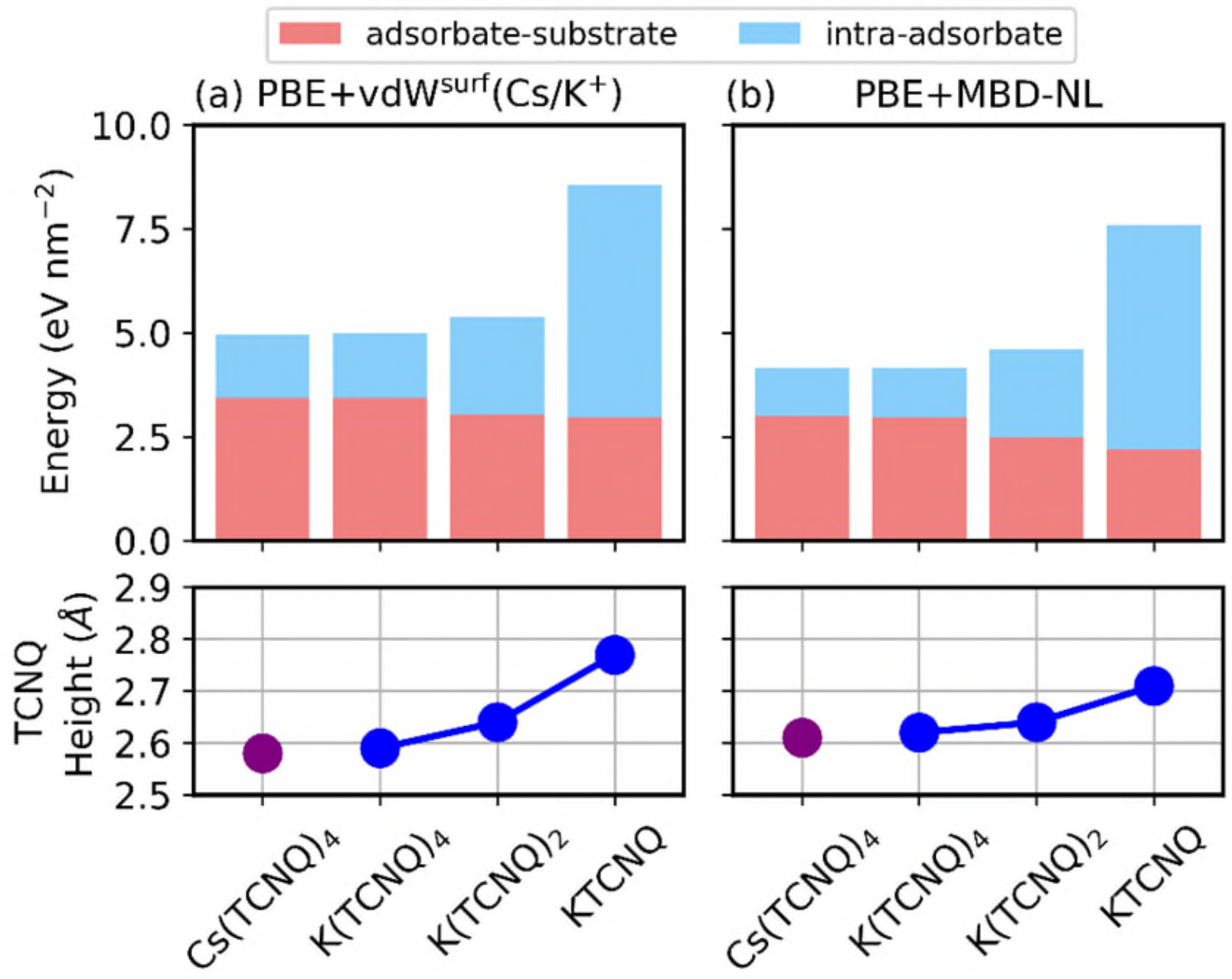

**Figure 7.** Bar plots showing energetic contributions of (light red) adsorbate to substrate and (light blue) intra-adsorbate interactions as determined by (a) PBE+vdW$^{surf}$(Cs/K+) and (b) PBE+MBD-NL. All values determined by Equation 1. (Bottom panels) Average height of C and N atoms in TCNQ molecules.

The stability of each network is also dependent upon the constituent stoichiometry of donor and acceptor. When changing the donor-acceptor ratio from 1:4, to 1:2 and to 1:1 for K-TCNQ adlayers, the total adsorption energy increases. As shown in Figure 7, this increase can be attributed mainly to the increase of cohesive interactions within the 2D-MOF layer. The increase corresponds to a doubling of this energy contribution from a stoichiometry of 1:4 to 1:2 (1.19 to 2.49 eV nm$^{-2}$ for MBD-NL) and a further doubling from 1:2 to 1:1 (2.49 to 5.41 eV nm$^{-2}$). This occurs at the slight expense of the adsorbate-substrate interaction, which is correlated with an increase in average adsorption height of the TCNQ molecule (bottom panel of Figure 7). Whereas the 2D-MOF cohesive energy accounts for only 29% of the total adsorption energy for the 1:4 alkali:TCNQ ratio, this increases to 54% for 1:2 and to 71% for 1:1. The increasing concentration of alkali atoms leads to a strengthening of the bonding within the 2D-MOF layer. At the same time, the adsorption height of the overlayer above the surface increases, which naturally leads to a slight reduction of the interaction between the 2D-MOF layer and the substrate. The findings are consistent with previous work where the presence of the alkali atoms was found to lead to a structural decoupling of TCNQ with an Ag(111) surface.[43] The observed stability trends are in qualitative agreement with experimental observations as the KTCNQ phase is created by annealing of other phases, which suggests that it is thermodynamically more stable.

Analysis of the charge distribution via Hirshfeld partitioning provides insight into the cause of the changes in interface stability (summarised in Tables S8-S11 in the ESI).[74] When the adsorbate-substrate interaction dominates, as in the case of $K(TCNQ)_4$, we find -0.75 *e* charge localised on each TCNQ molecule; the total net charge transferred from the surface to the 2D-MOF in this phase is -2.64 *e* per unit mesh, almost entirely accounting for the charge on the molecules. In the case of KTCNQ, in which the 2D-MOF cohesive energy dominates, we only find -0.32 *e*/TCNQ and almost no charge transfer from the surface (0.01 *e* per unit surface mesh). In $K(TCNQ)_2$, in which the energetic contributions are more equal, more charge transfer to the molecules (-0.91 *e*/TCNQ) is observed, in part from the substrate (-1.42 *e* per unit mesh), but in part from transfer within the 2D MOF. Thus, in the 1:4 and 1:2 stoichiometry phases, the surface acts as an electron donor leading to stronger adsorbate-surface interactions, but in the 1:1 alkali:TCNQ phase, the net charge transfer between the 2D-MOF and the surface is effectively zero. This leads to a decoupling of the 2D-MOF from the surface, whereas the charge balance between TCNQ and the alkali leads to strong ionic bonding within the 2D-MOF. Note that Hirshfeld partitioning is well known to underestimate the charge population on ions. While it is clear from density difference analysis that alkali atoms have a net charge of +1 *e* and are present within the 2D-MOF as cations, Hirshfeld partitioning only predicts a charge of between +0.3 to +0.5 *e*.

**Effect of overlayer composition on surface properties**

Adsorbates that are strong electron acceptors (donors) are well-known to increase (decrease) the work function at the interface due to charge transfer.[75–77] The overall change in work function $\Delta\phi$ due to adsorption of a molecular overlayer can be decomposed into two terms:

$$\Delta\phi = \Delta E_{\mathrm{bond}} + \Delta E_{\mathrm{MOF}} \qquad (2)$$

This includes a contribution due to the dipole density of the 2D MOF overlayer, $\Delta E_{\mathrm{MOF}}$, and a contribution due to the charge transfer and hybridisation of the adsorbate overlayer with the substrate, ΔEbond. We calculate $\Delta \mathrm{E}_{\mathrm{MOF}}$ as the potential drop through the free-stading 2D-MOF and evaluate ΔEbond as the difference between Δϕ and $\Delta \mathrm{E}_{\mathrm{MOF}}$. We refer to the second term as ΔEMOF as it arises from the dipole moment perpendicular to the surface of the complete 2D MOF rather than just the TCNQ molecules. Figure 8 shows the resulting changes in work function, together with these two separate contributions, relative to that of the clean Ag(100) surface as predicted by PBE+MBD-NL (4.19 eV).

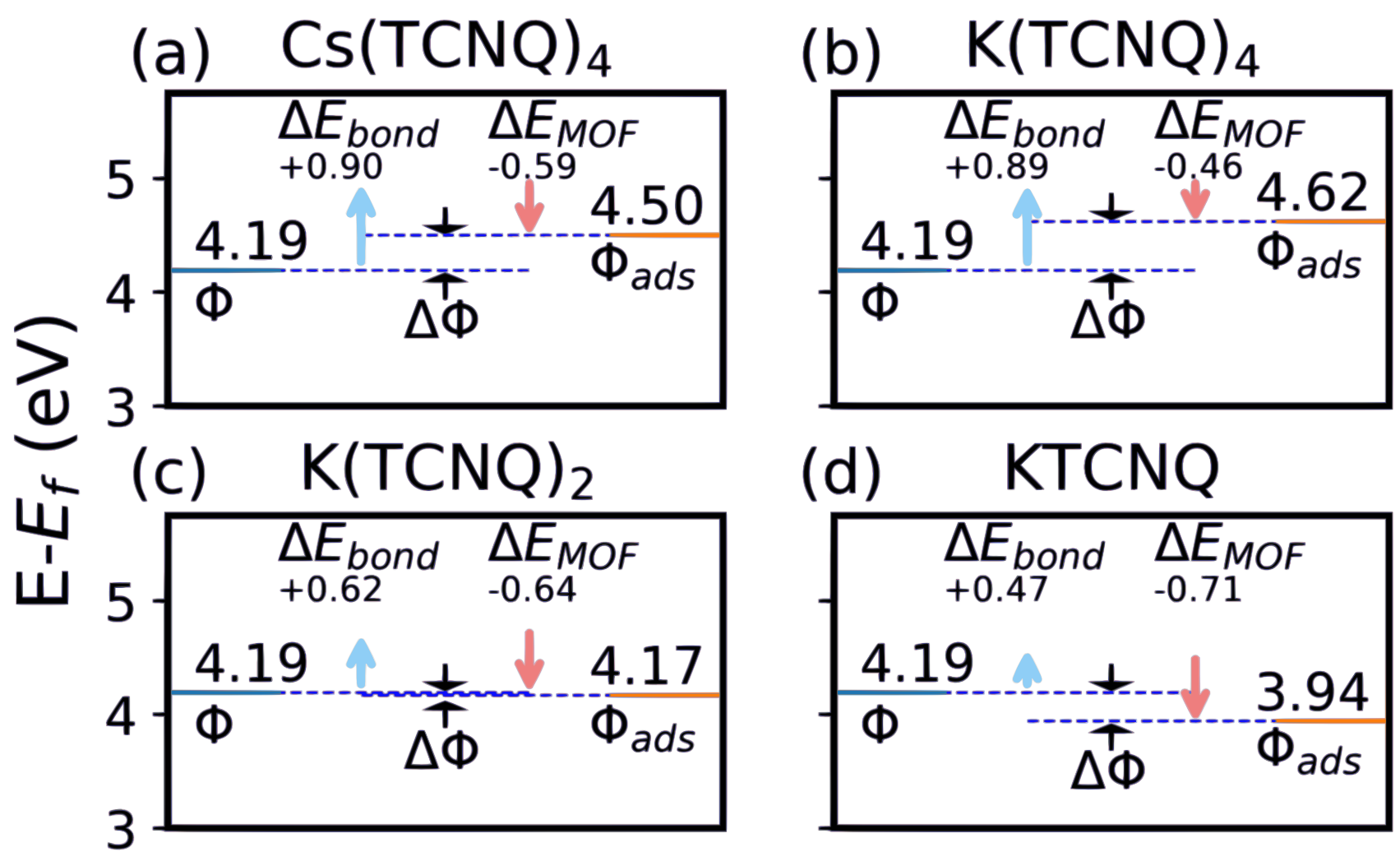

**Figure 8.** Graphical representation of the results of the theoretical work function analysis for (a) $Cs(TCNQ)_4$, (b) $K(TCNQ)_4$, (c) $K(TCNQ)_2$, and (d) KTCNQ. The work function of the clean surface, $\phi$, $\Delta E_{bond}$(light blue), $\Delta E_{MOF}$ (light red), as defined in equation 2, the work function after adsorption, $\phi_{ads}$ and the change in work function $\Delta\Phi$. All values calculated at the PBE+MBD-NL level.

For the 1:4 stoichiometry phases, with four electron-accepting TCNQ molecules and only one electron-donating alkali atom, our PBE+MBD-NL calculations show, as might be expected, that relative to the clean Ag(100) surface there are net increases in the work function (+0.31 eV and +0.43 eV for $Cs(TCNQ)_4$ and $K(TCNQ)_4$, respectively). However, increasing the fraction of alkali atoms to obtain the $KTCNQ_2$ phase leads to a small decrease in the work function (-0.02 eV), while in KTCNQ there is a significant work function decrease (to -0.25 eV). (see also Figure S4 for PBE+vdW$^{surf}$(Cs/K+) results and Tables S6 and S7 in the ESI for tabulated values). By going from a high TCNQ concentration to a 1:1 donor:acceptor ratio, the initial increase in work function upon adsorption ultimately changes to a decrease as dipole effects overwhelm chemical bonding effects.

$\Delta E_{\text{bond}}$ is essentially equal for $K(TCNQ)_4$ and $Cs(TCNQ)_4$ because the molecular heights are similar in these two structures, as is the charge transfer predicted by Hirshfeld analysis to the TCNQ ($Cs(TCNQ)_4$; -0.80 *e*/TCNQ, $K(TCNQ)_4$; -0.75 *e*/TCNQ, see Tables S8 and S9). The main difference in $\Delta\phi$ for these two phases is in the contribution due to the potential drop $\Delta E_{\text{MOF}}$, as $Cs(TCNQ)_4$ has a larger dipole moment than $K(TCNQ)_4$. This can be attributed to the fact that Cs is more spatially separated from the molecules in the layer than is K.

## 4. Conclusions

We report the formation and characterisation of several two-dimensional metal-organic frameworks composed of electron donating alkali atoms and the strong electron acceptor molecule, TCNQ, at various donor:acceptor ratios on Ag(100). We determine the adsorption structures via the quantitative experimental technique of NIXSW, complemented by state-of-the-art dispersion-inclusive density functional calculations of the experimentally identified adsorption phases with compositions $Cs(TCNQ)_4$, $K(TCNQ)_4$, $K(TCNQ)_2$ and KTCNQ. In all cases for which NIXSW data was available, we present structural models predicted by DFT that are in good agreement with experimental measurements. Whereas PBE+vdW$^{\text{surf}}$ with rescaled $C_6$ coefficients to account for the cationic nature of alkali atoms, vdW$^{\text{surf}}$(Cs/$K^+$), provides the most accurate adsorption height predictions, we find that the MBD-NL dispersion correction is not far off from vdW$^{\text{surf}}$(Cs/$K^+$) without the need for manual intervention. This shows that PBE+MBD-NL is able to provide robust structure predictions even in the case of strongly charge-separated systems with the additional benefit of mitigating the well-known overbinding of vdW$^{\text{surf}}$.

Having arrived at accurate structural models, we used DFT calculations to assess the stability of each phase. The $Cs(TCNQ)_4$ and $K(TCNQ)_4$ phases show similar energetic stability. In general, the nature and size of the alkali atom does not affect the cohesion of the two-dimensional metal-organic framework that is formed, although Cs, due to its larger size, is found at an elevated height above the plane of the molecular overlayer. By comparing K:TCNQ phases with different donor:acceptor ratios, we find trends in overall stability of the phases as well as the energetic contributions that arise from adlayer-substrate interaction and from the cohesion within the 2D-MOF. The overall stability of the network increases as the donor concentration is increased, mainly due to an increase in the strength of the ionic cohesive stabilisation within the layer. With a 1:1 alkali:TCNQ stoichiometry, the adlayer height increases and the 2D-MOF becomes more decoupled from the substrate than in the 1:4 case. Formation of the 1:4 phases ($K(TCNQ)_4$ and $Cs(TCNQ)_4$) leads to a strong increase in work function relative to the clean Ag(100) surface, 1:2 leaves the work function virtually unaffected, whereas the 1:1 phase reduces the work function of the Ag(100) surface.

In the context of device design, the results of this study can point to potential routes to exert control over observable properties at the interface via controlled sequential adsorption of alkali atoms and strong acceptor molecules to achieve varying donor:acceptor stoichiometries at the surface. The variation in structural composition has effects on the interface stability and bonding that are mediated by two-dimensional salt formation within the 2D-MOF layer. As a function of donor:acceptor ratio, surface electronic properties are strongly modified as dipole formation and molecular charge transfer vary due to competing interactions between alkali atoms, the substrate, and the acceptor molecule. At lower ratios (1:4) we find characteristic surface bound adlayers and at the donor acceptor parity (1:1), we find an organic charge transfer salt layer that is more weakly bonded to the surface. In the latter case, the surface dipole is mostly governed by the potential drop across this salt layer. As alkali atom deposition is an experimentally viable technique in thin-film manufacturing, our results show the significant structural and electronic changes that can be effected by controlled alkali atom doping.

**Supporting Information.** Additional detailed experimental results, experimental heights, additional height data from calculations, theoretical work function and charge redistribution data, molecular orbital density of states data.

**Acknowledgements**

The authors thank Diamond Light Source for allocations SI17261 and SI20785 of beam time at beamline I09 that contributed to the results presented here. P.T.P.R. and P.J.B. acknowledge financial support from Diamond Light Source and EPSRC. G.C. acknowledges financial support from the EU through the ERC Grant "VISUAL-MS" (Project ID: 308115). B.S. and R.J.M. acknowledge doctoral studentship funding from the EPSRC and the National Productivity Investment Fund (NPIF). R.J.M. acknowledges financial support via a UKRI Future Leaders Fellowship (MR/S016023/1). We acknowledge computing resources provided by the EPSRC-funded HPC Midlands+ Computing Centre (EP/P020232/1, EP/T022108/1) and the EPSRC-funded Materials Chemistry Consortium (EP/R029431/1) for the ARCHER2 U.K. National Supercomputing Service (http://www.archer2.ac.uk).

## References

(1) Braun, S.; Salaneck, W. R.; Fahlman, M. Energy-Level Alignment at Organic/Metal and Organic/Organic Interfaces. *Adv. Mater.* **2009**, *21*, 1450–1472.

(2) Schwartz, G.; Fehse, K.; Pfeiffer, M.; Walzer, K.; Leo, K. Highly Efficient White Organic Light Emitting Diodes Comprising an Interlayer to Separate Fluorescent and Phosphorescent Regions. *Appl. Phys. Lett.* **2006**, *89*, 083509.

(3) Pogantsch, A.; Rentenberger, S.; Langer, G.; Keplinger, J.; Kern, W.; Zojer, E. Tuning the Electroluminescence Color in Polymer Light-Emitting Devices Using the Thiol-Ene Photoreaction. *Adv. Funct. Mater.* **2005**, *15*, 403–409.

(4) Ma, W.; Yang, C.; Gong, X.; Lee, K.; Heeger, A. J. Thermally Stable, Efficient Polymer Solar Cells with Nanoscale Control of the Interpenetrating Network Morphology. *Adv. Funct. Mater.* **2005**, *15*, 1617–1622.

(5) Hoppe, H.; Sariciftci, N. S. Morphology of Polymer/Fullerene Bulk Heterojunction Solar Cells. *J. Mater. Chem.* **2006**, *16*, 45–61.

(6) Crispin, X.; Geskin, V.; Crispin, A.; Cornil, J.; Lazzaroni, R.; Salaneck, W. R.; Brédas, J.-L. Characterization of the Interface Dipole at Organic/ Metal Interfaces. *J. Am. Chem. Soc.* **2002**, *124*, 8131–8141.

(7) Bagus, P. S.; Staemmler, V.; Wöll, C. Exchangelike Effects for Closed-Shell Adsorbates: Interface Dipole and Work Function. *Phys. Rev. Lett.* **2002**, *89*, 096104.

(8) Witte, G.; Lukas, S.; Bagus, P. S.; Wöll, C. Vacuum Level Alignment at Organic/Metal Junctions: “Cushion” Effect and the Interface Dipole. *Appl. Phys. Lett.* **2005**, *87*, 263502.

(9) Bagus, P. S.; Hermann, K.; Wöll, C. The Interaction of C6H6 and C6H12 with Noble Metal Surfaces: Electronic Level Alignment and the Origin of the Interface Dipole. *The Journal of Chemical Physics* **2005**, *123*, 184109.

(10) Koch, N.; Kahn, A.; Ghijsen, J.; Pireaux, J.-J.; Schwartz, J.; Johnson, R. L.; Elschner, A. Conjugated Organic Molecules on Metal versus Polymer Electrodes: Demonstration of a Key Energy Level Alignment Mechanism. *Appl. Phys. Lett.* **2003**, *82*, 70–72.

(11) Zojer, E.; Taucher, T. C.; Hofmann, O. T. The Impact of Dipolar Layers on the Electronic Properties of Organic/Inorganic Hybrid Interfaces. *Adv. Mater. Interfaces* **2019**, *6*, 1900581.

(12) Zuppiroli, L.; Si-Ahmed, L.; Kamaras, K.; Nüesch, F.; Bussac, M. N.; Ades, D.; Siove, A.; Moons, E.; Grätzel, M. Self-Assembled Monolayers as Interfaces for Organic Opto-Electronic Devices. *Eur. Phys. J. B* **1999**, *11*, 505–512.

(13) Schulz, P.; Schäfer, T.; Zangmeister, C. D.; Effertz, C.; Meyer, D.; Mokros, D.; van Zee, R. D.; Mazzarello, R.; Wuttig, M. A New Route to Low Resistance Contacts for Performance-Enhanced Organic Electronic Devices. *Adv. Mater. Interfaces* **2014**, *1*, 1300130.

(14) Casalini, S.; Bortolotti, C. A.; Leonardi, F.; Biscarini, F. Self-Assembled Monolayers in Organic Electronics. *Chem. Soc. Rev.* **2017**, *46*, 40–71.

(15) Gonzalez-Lakunza, N.; Fernández-Torrente, I.; Franke, K. J.; Lorente, N.; Arnau, A.; Pascual, J. I. Formation of Dispersive Hybrid Bands at an Organic-Metal Interface. *Phys. Rev. Lett.* **2008**, *100*, 156805.
(16) de Oteyza, D. G.; GarcÃ-a-Lastra, J. M.; Corso, M.; Doyle, B. P.; Floreano, L.; Morgante, A.; Wakayama, Y.; Rubio, A.; Ortega, J. E. Customized Electronic Coupling in Self-Assembled Donor–Acceptor Nanostructures. *Adv. Funct. Mater.* **2009**, *19*, 3567–3573.
(17) Goiri, E.; Matena, M.; El-Sayed, A.; Lobo-Checa, J.; Borghetti, P.; Rogero, C.; Detlefs, B.; Duvernay, J.; Ortega, J. E.; de Oteyza, D. G. Self-Assembly of Bicomponent Molecular Monolayers: Adsorption Height Changes and Their Consequences. *Phys. Rev. Lett.* **2014**, *112*, 117602.
(18) Umbach, T. R.; Fernandez-Torrente, I.; Ladenthin, J. N.; Pascual, J. I.; Franke, K. J. Enhanced Charge Transfer in a Monolayer of the Organic Charge Transfer Complex TTF–TNAP on Au(111). *J. Phys.: Condens. Matter* **2012**, *24*, 354003.
(19) Erley, W.; Ibach, H. Vibrational Spectra of Tetracyanoquinodimethane (TCNQ) Adsorbed on the Cu(111) Surface. *Surface Science* **1986**, *178*, 565–577.
(20) Barja, S.; Stradi, D.; Borca, B.; Garnica, M.; Díaz, C.; Rodriguez-García, J. M.; Alcamí, M.; Vázquez de Parga, A. L.; Martín, F.; Miranda, R. Ordered Arrays of Metal–Organic Magnets at Surfaces. *J. Phys.: Condens. Matter* **2013**, *25*, 484007.
(21) Giergiel, J.; Wells, S.; Land, T. A.; Hemminger, J. C. Growth and Chemistry of TCNQ Films on Nickel (111). *Surface Science* **1991**, *255*, 31–40.
(22) Feyer, V.; Graus, M.; Nigge, P.; Zamborlini, G.; Acres, R. G.; Schöll, A.; Reinert, F.; Schneider, C. M. The Geometric and Electronic Structure of TCNQ and TCNQ+Mn on Ag(0 0 1) and Cu(0 0 1) Surfaces. *Journal of Electron Spectroscopy and Related Phenomena* **2015**, *204*, 125–131.
(23) Deilmann, T.; Krüger, P.; Rohlfing, M.; Wegner, D. Adsorption and STM Imaging of Tetracyanoethylene on Ag(001): An *Ab Initio* Study. *Phys. Rev. B* **2014**, *89*, 045405.
(24) Wegner, D.; Yamachika, R.; Wang, Y.; Brar, V. W.; Bartlett, B. M.; Long, J. R.; Crommie, M. F. Single-Molecule Charge Transfer and Bonding at an Organic/Inorganic Interface: Tetracyanoethylene on Noble Metals. *Nano Lett.* **2008**, *8*, 131–135.
(25) Koch, N. Energy Levels at Interfaces between Metals and Conjugated Organic Molecules. *J. Phys.: Condens. Matter* **2008**, *20*, 184008.
(26) Akaike, K.; Koch, N.; Heimel, G.; Oehzelt, M. The Impact of Disorder on the Energy Level Alignment at Molecular Donor–Acceptor Interfaces. *Advanced Materials Interfaces* **2015**, *2*, 1500232.
(27) Sun, Z.; Shi, S.; Bao, Q.; Liu, X.; Fahlman, M. Role of Thick-Lithium Fluoride Layer in Energy Level Alignment at Organic/Metal Interface: Unifying Effect on High Metallic Work Functions. *Advanced Materials Interfaces* **2015**, *2*, 1400527.
(28) Cartus, J. J.; Jeindl, A.; Hofmann, O. T. Can We Predict Interface Dipoles Based on Molecular Properties? *ACS Omega* **2021**, *6*, 32270–32276.
(29) Shan, H.; Zhou, L.; Ji, W.; Zhao, A. Flexible Alkali–Halogen Bonding in Two Dimensional Alkali-Metal Organic Frameworks. *J. Phys. Chem. Lett.* **2021**, *12*, 10808–10814.
(30) Stepanow, S.; Ohmann, R.; Leroy, F.; Lin, N.; Strunskus, T.; Wöll, C.; Kern, K. Rational Design of Two-Dimensional Nanoscale Networks by Electrostatic Interactions at Surfaces. *ACS Nano* **2010**, *4*, 1813–1820.
(31) Fahlman, M.; Crispin, A.; Crispin, X.; Henze, S. K. M.; Jong, M. P. de; Osikowicz, W.; Tengstedt, C.; Salaneck, W. R. Electronic Structure of Hybrid Interfaces for Polymer-Based Electronics. *J. Phys.: Condens. Matter* **2007**, *19*, 183202.
(32) Zwick, C.; Baby, A.; Gruenewald, M.; Verwüster, E.; Hofmann, O. T.; Forker, R.; Fratesi, G.; Brivio, G. P.; Zojer, E.; Fritz, T. Complex Stoichiometry-Dependent Reordering of 3,4,9,10-Perylenetetracarboxylic Dianhydride on Ag(111) upon K Intercalation. *ACS Nano* **2016**, *10*, 2365–2374.
(33) Baby, A.; Gruenewald, M.; Zwick, C.; Otto, F.; Forker, R.; van Straaten, G.; Franke, M.; Stadtmüller, B.; Kumpf, C.; Brivio, G. P.; et al. Fully Atomistic Understanding of the Electronic and Optical Properties of a Prototypical Doped Charge-Transfer Interface. *ACS Nano* **2017**, *11*, 10495–10508.
(34) Chi, X.; Besnard, C.; Thorsmølle, V. K.; Butko, V. Y.; Taylor, A. J.; Siegrist, T.; Ramirez, A. P. Structure and Transport Properties of the Charge-Transfer Salt Coronene−TCNQ. *Chem. Mater.* **2004**, *16*, 5751–5755.
(35) Han, B.; Ma, R.; Wang, H.; Zhou, M. High Pressure Investigations on TTF-TCNQ Charge-Transfer Complexes. *Spectrochimica Acta Part A: Molecular and Biomolecular Spectroscopy* **2022**, *267*, 120541.
(36) Kirtley, J. R.; Mannhart, J. When TTF Met TCNQ. *Nature Mater* **2008**, *7*, 520–521.
(37) Hsu, C.-L.; Lin, C.-T.; Huang, J.-H.; Chu, C.-W.; Wei, K.-H.; Li, L.-J. Layer-by-Layer Graphene/TCNQ Stacked Films as Conducting Anodes for Organic Solar Cells. *ACS Nano* **2012**, *6*, 5031–5039.
(38) Ferraris, J.; Cowan, D. O.; Walatka, V.; Perlstein, J. H. *Electron transfer in a new highly conducting donor-acceptor complex.* ACS Publications. https://pubs.acs.org/doi/pdf/10.1021/ja00784a066 (accessed 2022-07-28).
(39) Blowey, P. J.; Velari, S.; Rochford, L. A.; Duncan, D. A.; Warr, D. A.; Lee, T.-L.; De Vita, A.; Costantini, G.; Woodruff, D. P. Re-Evaluating How Charge Transfer Modifies the Conformation of Adsorbed Molecules. *Nanoscale* **2018**, *10*, 14984–14992.
(40) Haags, A.; Rochford, L. A.; Felter, J.; Blowey, P. J.; Duncan, D. A.; Woodruff, D. P.; Kumpf, C. Growth and Evolution of Tetracyanoquinodimethane and Potassium Coadsorption Phases on Ag(111). *New J. Phys.* **2020**, *22*, 063028.
(41) Ryan, P.; Blowey, P. J.; Sohail, B. S.; Rochford, L. A.; Duncan, D. A.; Lee, T.-L.; Starrs, P.; Costantini, G.; Maurer, R. J.; Woodruff, D. P. Thermodynamic Driving Forces for Substrate Atom Extraction by Adsorption of Strong Electron Acceptor Molecules. *J. Phys. Chem. C* **2022**, *126*, 6082–6090.
(42) Floris, A.; Comisso, A.; De Vita, A. Fine-Tuning the Electrostatic Properties of an Alkali-Linked Organic Adlayer on a Metal Substrate. *ACS Nano* **2013**, *7*, 8059–8065.
(43) Blowey, P. J.; Sohail, B.; Rochford, L. A.; Lafosse, T.; Duncan, D. A.; Ryan, P. T. P.; Warr, D. A.; Lee, T.-L.; Costantini, G.; Maurer, R. J.; et al. Alkali Doping Leads to Charge-Transfer Salt Formation in a Two-Dimensional Metal–Organic Framework. *ACS Nano* **2020**, *14*, 7475–7483.
(44) Hofmann, O. T.; Zojer, E.; Hörmann, L.; Jeindl, A.; Maurer, R. J. First-Principles Calculations of Hybrid Inorganic–Organic Interfaces: From State-of-the-Art to Best Practice. *Phys. Chem. Chem. Phys.* **2021**, *23*, 8132–8180.
(45) Maurer, R. J.; Ruiz, V. G.; Camarillo-Cisneros, J.; Liu, W.; Ferri, N.; Reuter, K.; Tkatchenko, A. Adsorption Structures and Energetics of Molecules on Metal Surfaces: Bridging Experiment and Theory. *Progress in Surface Science* **2016**, *91*, 72–100.
(46) Maurer, R. J.; Ruiz, V. G.; Tkatchenko, A. Many-Body Dispersion Effects in the Binding of Adsorbates on Metal Surfaces. *The Journal of Chemical Physics* **2015**, *143*, 102808.
(47) Carrasco, J.; Liu, W.; Michaelides, A.; Tkatchenko, A. Insight into the Description of van Der Waals Forces for Benzene Adsorption on Transition Metal (111) Surfaces. *The Journal of Chemical Physics* **2014**, *140*, 084704.
(48) Gould, T.; Bučko, T. $C_6$ Coefficients and Dipole Polarizabilities for All Atoms and Many Ions in Rows 1–6 of the Periodic Table. *J. Chem. Theory Comput.* **2016**, *12*, 3603–3613.

(49) Hermann, J.; Tkatchenko, A. Density Functional Model for van Der Waals Interactions: Unifying Many-Body Atomic Approaches with Nonlocal Functionals. *Phys. Rev. Lett.* **2020**, *124*, 146401.
(50) Abdurakhmanova, N.; Floris, A.; Tseng, T.-C.; Comisso, A.; Stepanow, S.; De Vita, A.; Kern, K. Stereoselectivity and Electrostatics in Charge-Transfer Mn- and Cs-TCNQ4 Networks on Ag(100). *Nat Commun* **2012**, *3*, 940.
(51) Lee, T.-L.; Duncan, D. A. A Two-Color Beamline for Electron Spectroscopies at Diamond Light Source. *Synchrotron Radiation News* **2018**, *31*, 16–22.
(52) Nečas, D.; Klapetek, P. Gwyddion: An Open-Source Software for SPM Data Analysis. *Open Physics* **2012**, *10*.
(53) Woodruff, D. P. Surface Structure Determination Using X-Ray Standing Waves. *Rep. Prog. Phys.* **2005**, *68*, 743–798.
(54) Nefedov, V. I.; Yarzhemsky, V. G.; Nefedova, I. S.; Trzhaskovskaya, M. B.; Band, I. M. The Influence of Non-Dipolar Transitions on the Angular Photoelectron Distribution. *Journal of Electron Spectroscopy and Related Phenomena* **2000**, *107*, 123–130.
(55) Blum, V.; Gehrke, R.; Hanke, F.; Havu, P.; Havu, V.; Ren, X.; Reuter, K.; Scheffler, M. Ab Initio Molecular Simulations with Numeric Atom-Centered Orbitals. *Computer Physics Communications* **2009**, *180*, 2175–2196.
(56) Perdew, J. P.; Burke, K.; Ernzerhof, M. Generalized Gradient Approximation Made Simple. *Physical review letters* **1996**, No. 3, 3865–3868.
(57) Tkatchenko, A.; Scheffler, M. Accurate Molecular van Der Waals Interactions from Ground-State Electron Density and Free-Atom Reference Data. *Physical Review Letters* **2009**, *102*, 6–9.
(58) Hermann, J.; Tkatchenko, A. Density Functional Model for van Der Waals Interactions : Unifying Many-Body Atomic Approaches with Nonlocal Functionals. *Physical Review Letters* **2020**, *124*, 146401.
(59) Ruiz, V. G.; Liu, W.; Zojer, E.; Scheffler, M.; Tkatchenko, A. Density-Functional Theory with Screened van Der Waals Interactions for the Modeling of Hybrid Inorganic-Organic Systems. *Phys. Rev. Lett.* **2012**, *108*, 146103.
(60) Monkhorst, H. J.; Pack, J. D. Special Points for Brillouin-Zone Integrations. *Physical Review B* **-** *Condensed Matter and Materials Physics* **1976**, *13*, 5188–5192.
(61) Faraggi, M. N.; Jiang, N.; Gonzalez-Lakunza, N.; Langner, A.; Stepanow, S.; Kern, K.; Arnau, A. Bonding and Charge Transfer in Metal–Organic Coordination Networks on Au(111) with Strong Acceptor Molecules. *J. Phys. Chem. C* **2012**, *116*, 24558–24565.
(62) Stradi, D.; Borca, B.; Barja, S.; Garnica, M.; Díaz, C.; Rodríguez-García, J. M.; Alcamí, M.; Vázquez de Parga, A. L.; Miranda, R.; Martín, F. Understanding the Self-Assembly of TCNQ on Cu(111): A Combined Study Based on Scanning Tunnelling Microscopy Experiments and Density Functional Theory Simulations. *RSC Adv.* **2016**, *6*, 15071–15079.
(63) Kamna, M. M.; Graham, T. M.; Love, J. C.; Weiss, P. S. Strong Electronic Perturbation of the Cu{111} Surface by 7,7′,8,8′-Tetracyanoquinonedimethane. *Surface Science* **1998**, *419*, 12–23.
(64) Blowey, P. J.; Sohail, B.; Rochford, L. A.; Lafosse, T.; Duncan, D. A.; Ryan, P. T. P.; Warr, D. A.; Lee, T.; Costantini, G.; Maurer, R. J.; et al. Alkali Doping Leads to Charge-Transfer Salt Formation in a Two-Dimensional Metal − Organic Framework. **2020**.
(65) Blowey, P. J.; Rochford, L. A.; Duncan, D. A.; Ryan, P. T. P.; Warr, D. A.; Lee, T.-L.; Costantini, G.; Woodruff, D. P. The Structure of 2D Charge Transfer Salts Formed by TCNQ/Alkali Metal Coadsorption on Ag(111). *Surface Science* **2020**, *701*, 121687.

(66) Blowey, P. J. Probing the Geometrical and Electronic Structure of Two Dimensional Charge Transfer Networks on Metal Surfaces, University of Warwick, 2018. http://webcat.warwick.ac.uk/record=b3253314~S15.
(67) Zegenhagen, J. Surface Structure Determination with X-Ray Standing Waves. *Surface Science Reports* **1993**, *18*, 202–271.
(68) Bedzyk, M. J.; Materlik, G. Two-Beam Dynamical Diffraction Solution of the Phase Problem: A Determination with x-Ray Standing-Wave Fields. *Phys. Rev. B* **1985**, *32*, 6456–6463.
(69) Woodruff, D. P.; Duncan, D. A. X-Ray Standing Wave Studies of Molecular Adsorption: Why Coherent Fractions Matter. *New J. Phys.* **2020**, *22*, 113012.
(70) Yeh, J. J.; Lindau, I. Atomic Subshell Photoionization Cross Sections and Asymmetry Parameters: $1 \leqslant Z \leqslant 103$. *Atomic Data and Nuclear Data Tables* **1985**, *32*, 1–155.
(71) Perdew, J. P.; Burke, K.; Ernzerhof, M. Generalized Gradient Approximation Made Simple. *Phys. Rev. Lett.* **1996**, *77*, 3865–3868.
(72) Blum, V.; Gehrke, R.; Hanke, F.; Havu, P.; Havu, V.; Ren, X.; Reuter, K.; Scheffler, M. Ab Initio Molecular Simulations with Numeric Atom-Centered Orbitals. *Computer Physics Communications* **2009**, *180*, 2175–2196.
(73) Maurer, R. J.; Freysoldt, C.; Reilly, A. M.; Brandenburg, J. G.; Hofmann, O. T.; Björkman, T.; Lebègue, S.; Tkatchenko, A. Advances in Density-Functional Calculations for Materials Modeling. *Annu. Rev. Mater. Res.* **2019**, *49*, 1–30.
(74) Hirshfeld, F. L. Bonded-Atom Fragments for Describing Molecular Charge Densities. *Theoret. Chim. Acta* **1977**, *44*, 129–138.
(75) Ertl, G.; Lee, S. B.; Weiss, M. Adsorption of Nitrogen on Potassium Promoted Fe(111) and (100) Surfaces. *Surface Science* **1982**, *114*, 527–545.
(76) Brodén, G.; Bonzel, H. P. Potassium Adsorption on Fe(110). *Surface Science* **1979**, *84*, 106–120.
(77) Paál, Z.; Ertl, G.; Lee, S. B. Interactions of Potassium, Oxygen and Nitrogen with Polycrystalline Iron Surfaces. *Applications of Surface Science* **1981**, *8*, 231–249.

**TOC Graphic**

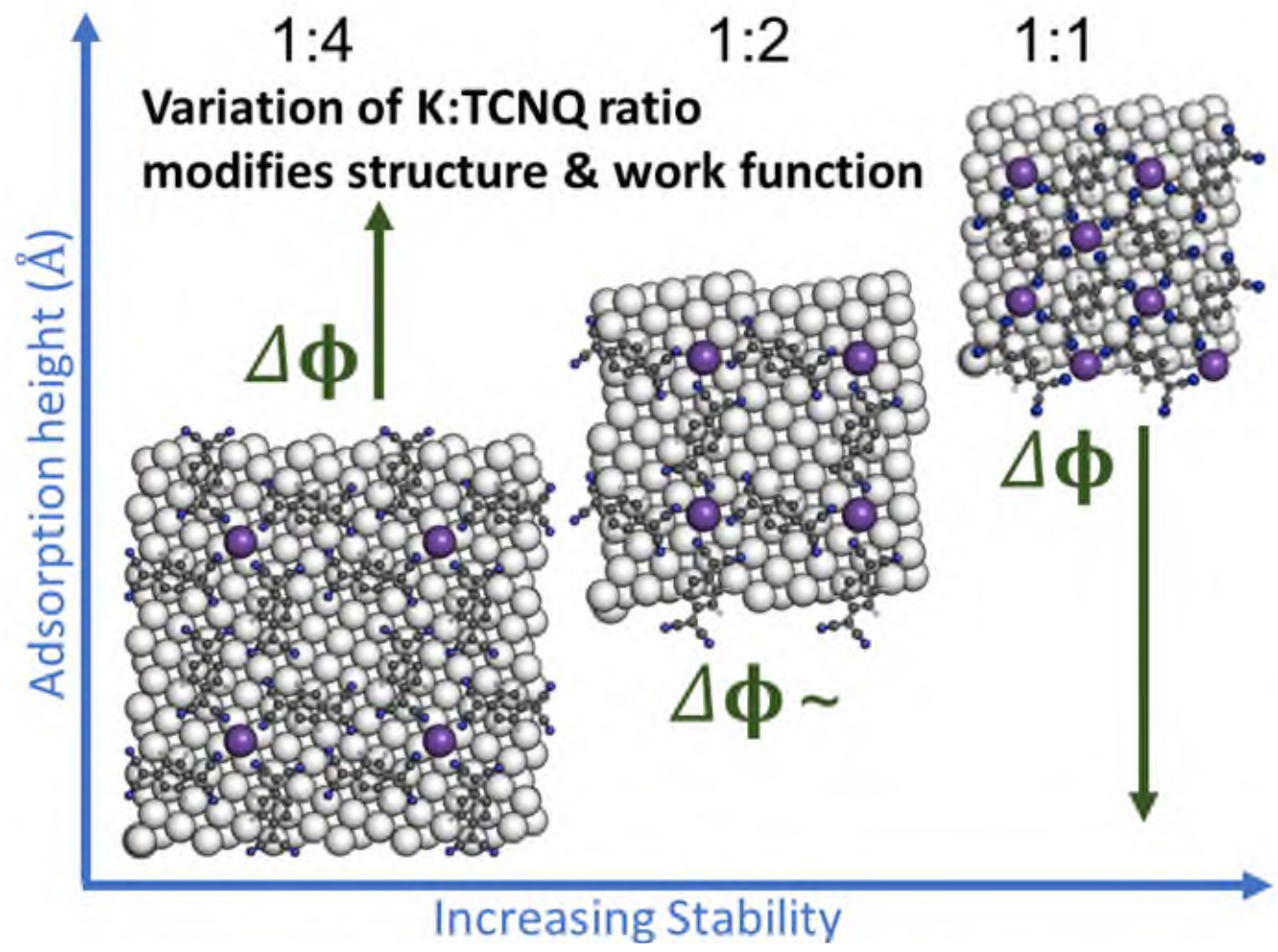

**Supporting Information to**

# Donor-Acceptor Co-Adsorption Ratio Controls Structure and Electronic Properties of Two-Dimensional Alkali-Organic Networks on Ag(100)

B. Sohail[1], P.J. Blowey[2,3], L.A. Rochford[4], P.T.P. Ryan[3,5], D.A. Duncan[3], T.-L. Lee[3], P. Starrs [3,6], G. Costantini[2,4], D.P. Woodruff[2*], R.J. Maurer[1,2*]

*(1) Department of Chemistry, University of Warwick, Coventry CV4 7AL, UK*

*(2) Department of Physics, University of Warwick, Coventry CV4 7AL, UK*

*(3) Diamond Light Source, Harwell Science and Innovation Campus, Didcot, OX11 0DE, UK*

*(4) School of Chemistry, University of Birmingham, Birmingham B15 2TT, UK*

*(5) Department of Materials, Imperial College, London SW7 2AZ, UK*

*(6) School of Chemistry, University of St. Andrews, St. Andrews, KY16 9AJ, UK*

## Contents

* Email: r.maurer@warwick.ac.uk, d.p.woodruff@warwick.ac.uk

# 1. Additional detailed experimental results: LEED, SXPS and NIXSW

LEED patterns from the 4 different alkali-TCNQ adsorption phases are shown in Figure S1.

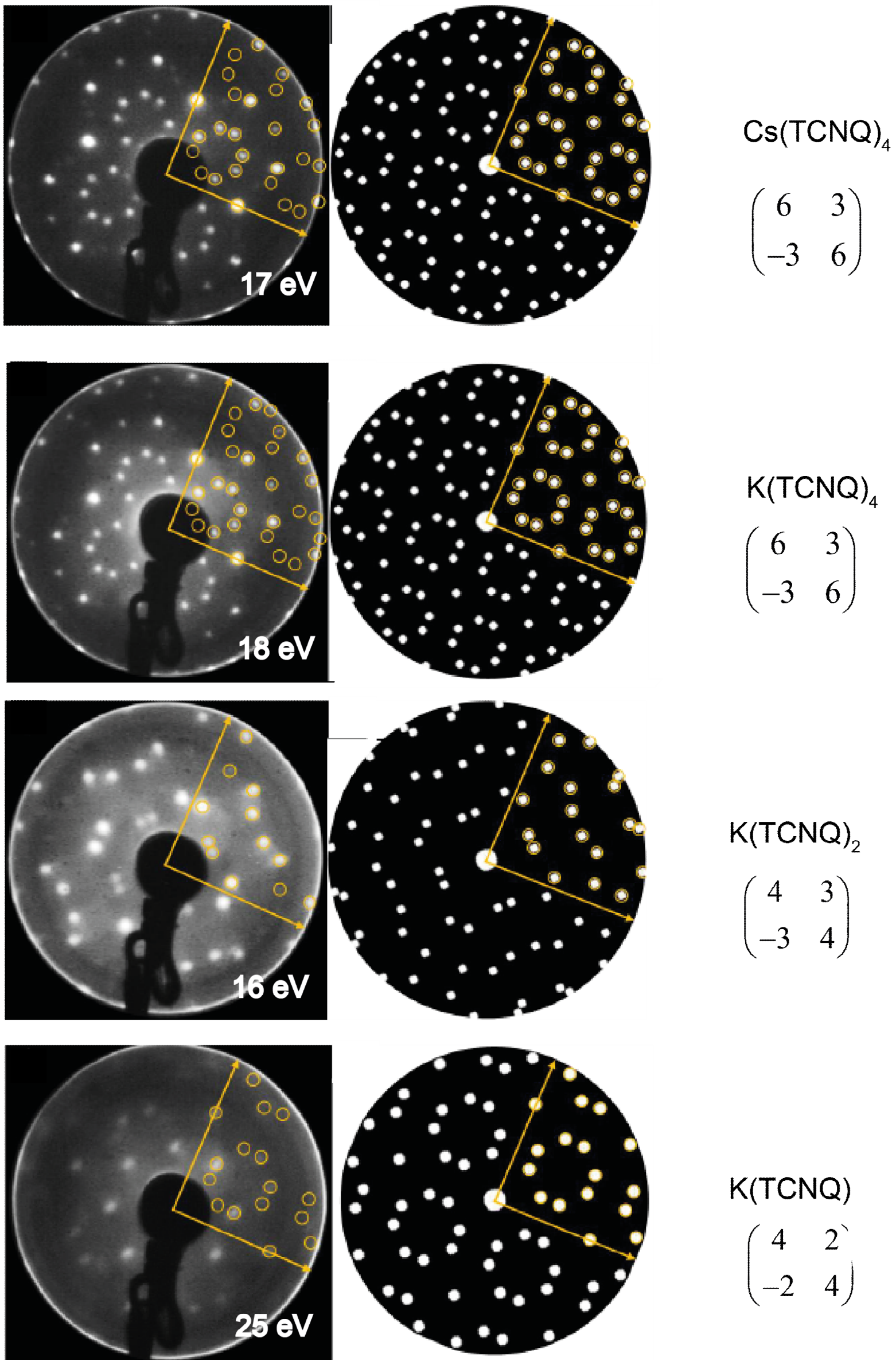

**Figure S1**. Experimental LEED patterns (left) and simulations of these patterns using the LEEDpat[1] program (right) for each of the 4 different alkali-TCNQ coadsorption phases. Circled in yellow are all the predicted beams in the upper-right hand quadrant, these circles also being repeated in the experimental LEED patterns. The yellow arrows show the <110> directions of the Ag(100) substrate.

A summary of the different phases found is given in Table S1

**Table S1** Summary of the different ordered co-adsorption phases of TCNQ and K/Cs found on Ag(100).

| **Phase descriptor** | **Matrix** | **Unit mesh area (Å²)** | **No of molecules per unit mesh** | **No of K/Cs atoms per unit mesh** | **Area per molecule (Å²)** | **Preparation** |
|---|---|---|---|---|---|---|
| $KTCNQ_4$ and $CsTCNQ_4$ | $\begin{pmatrix} 6 & 3 \\ -3 & 6 \end{pmatrix}$ | 375 | 4 | 1 | 94 | Alkali deposition onto Ag(100)$\begin{pmatrix} 1 & 4 \\ -3 & -1 \end{pmatrix}$- TCNQ surface |
| $KTCNQ_2$ | $\begin{pmatrix} 4 & 3 \\ -3 & 4 \end{pmatrix}$ | 210 | 2 | 1 | 105 | Additional K deposition onto $KTCNQ_4$ phase |
| KTCNQ | $\begin{pmatrix} 4 & 2 \\ -2 & 4 \end{pmatrix}$ | 166 | 2 | 2 | 83 | Additional K deposition onto $KTCNQ_2$ and annealing to ~300°C |

SXPS data from the $K(TCNQ)_4$ phase are shown in Figure 2 of the main manuscript, while SXPS from the $K(TCNQ)_2$ and $Cs(TCNQ)_4$ phases are shown in Figure S2

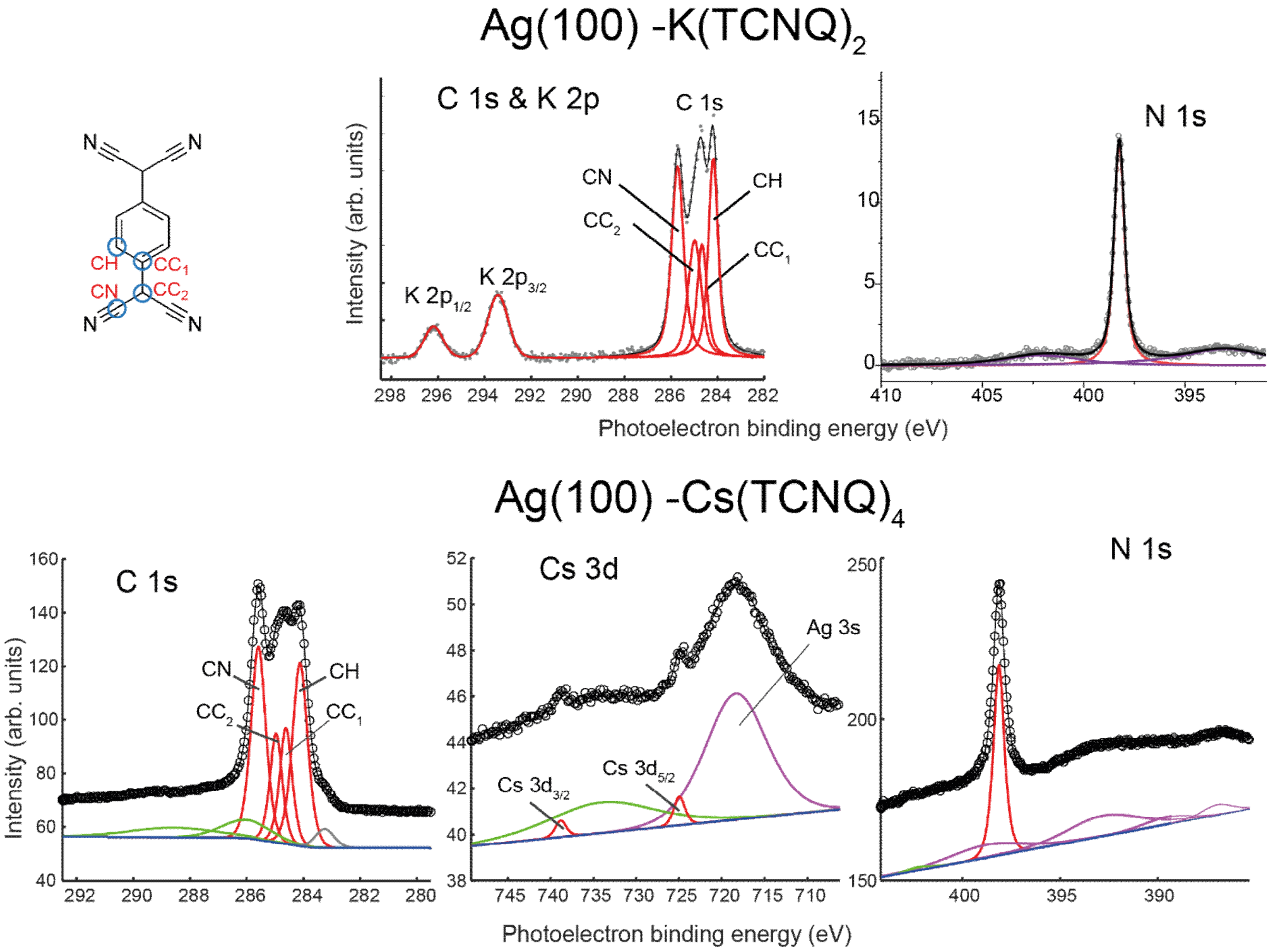

**Figure S2** C 1s and K 2p, Cs 3d, and N 1s SXP spectra recorded from the K(TCNQ)$_2$ and Cs(TCNQ)$_4$ phases on Ag(100) at photon energies of 435 eV, 900 eV and 550 eV respectively. The main photoemission peaks (including the different chemically-shifted C 1s peaks) are shown in red. Satellites are shown in green while the plasmon satellites of the Ag 3d emission in the N 1s spectrum are shown in purple. A schematic of the TCNQ molecule shows the labelling of the inequivalent C atoms that are distinguished in the C 1s spectra.

Figure S3 shows a comparison of the raw NIXSW photoemission intensity scans from the three alakli/TCNQ coadsorption phases investigated by this technique with the best-fit theoretical curves, thecorresponding values of the two fitting parameters (coherent fraction and coherent position) being reported in Table 1 of the main paper. The analysis included corrections for the backward/forward asymmetry of the photoemission angular dependence.

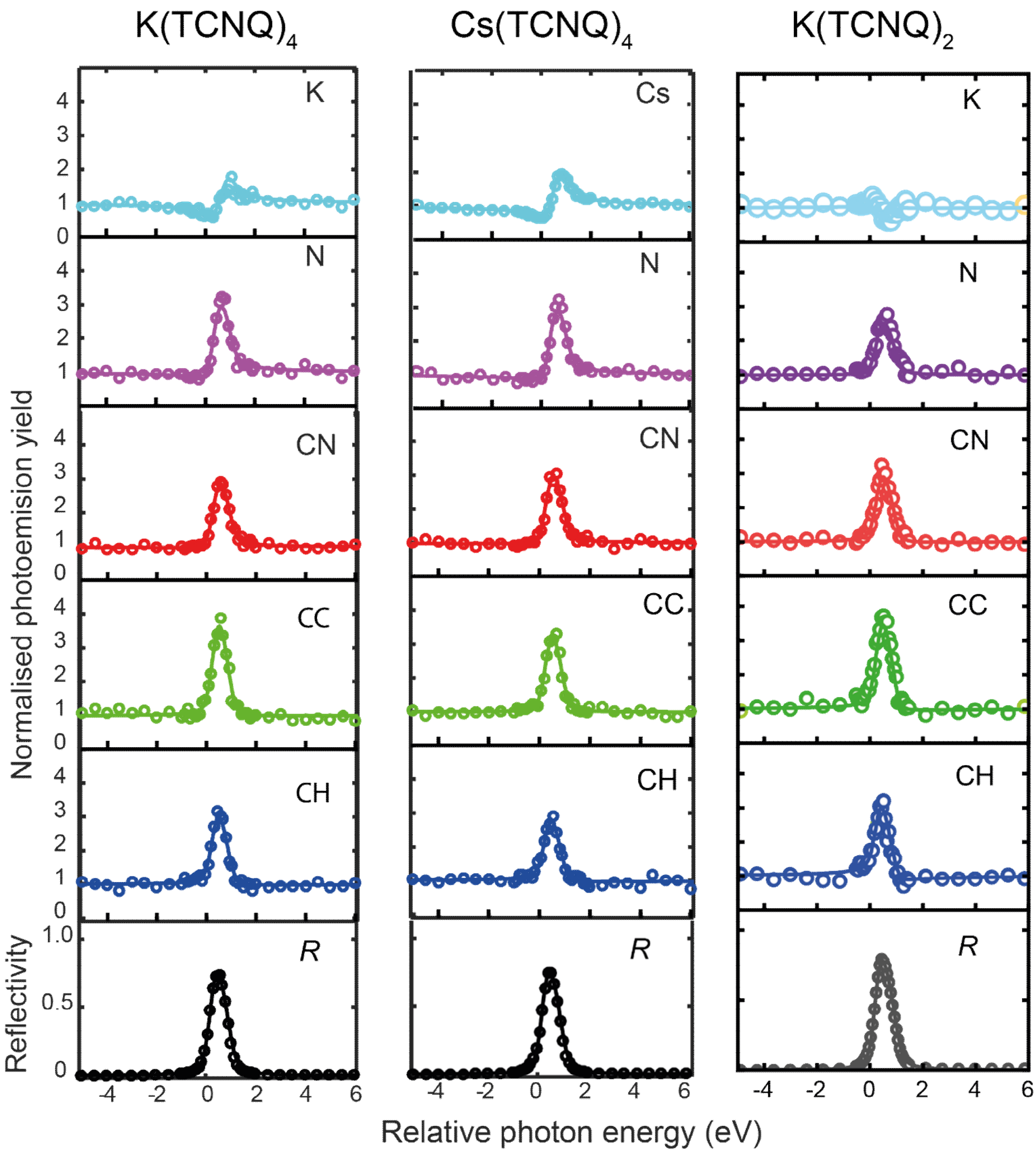

**Figure S3** Chemical-state specific NIXSW photoemission data for the (200) Bragg reflection from the $K(TCNQ)_4$, $Cs(TCNQ)_4$ and $K(TCNQ)_2$ surface phases. Experimental data points are shown as circles while the continuous lines are theoretical fits corresponding to the coherent fraction and position values reported in Table 1.

## 2. Comparison of predicted NIXSW structural parameters using different dispersion schemes

Tables S2-S5 provide a comparison of the experimental NIXSW structural parameters and the predicted values of these parameters determined from DFT calculations for 3 dispersion schemes employed, namely PBE+vdW$^{surf}$(Cs/K+),[2] PBE+vdW$^{surf}$,[3] and PBE+MBD-NL[4]. Notice that the theoretical values of the coherent fractions, *f*, however, take no account of the static or dynamic disorder, which may be present in the experimental values.

**Table S2** Comparison of the experiential NIXSW structural parameters for the Ag(100) $Cs(TCNQ)_4$ phase with the predictions of the three different dispersion schemes.

| | ***f*** **expt** | ***D*** **(Å) expt** | ***f*** **vdW$^{surf}$-Cs$^+$** | ***D*** **(Å) vdW$^{surf}$-Cs$^+$** | ***f*** **vdW$^{surf}$** | ***D*** **(Å) vdW$^{surf}$** | ***f*** **MBD-NL** | ***D*** **(Å) MBD-NL** |
|---|---|---|---|---|---|---|---|---|
| C-H | 0.63(10) | 2.72(5) | 0.99 | 2.69 | 0.99 | 2.70 | 1.00 | 2.75 |
| C-C | 0.80(10) | 2.64(5) | 0.99 | 2.64 | 0.99 | 2.64 | 0.98 | 2.69 |
| C-N | 0.71(10) | 2.53(5) | 0.97 | 2.55 | 0.98 | 2.54 | 0.90 | 2.60 |
| N | 0.78(10) | 2.38(5) | 0.90 | 2.45 | 0.93 | 2.43 | 0.71 | 2.46 |
| Cs | 0.76(10) | 4.08(5) | 1.00 | 4.06 | 1.00 | 2.86 | 1.00 | 3.76 |

**Table S3** Comparison of the experiential NIXSW structural parameters for the Ag(100) $K(TCNQ)_4$ phase with the predictions of the three different dispersion schemes.

| | ***f*** **expt** | ***D*** **(Å) expt** | ***f*** **vdW$^{surf}$-K$^+$** | ***D*** **(Å) vdW$^{surf}$-K$^+$** | ***f*** **vdW$^{surf}$** | ***D*** **(Å) vdW$^{surf}$** | ***f*** **MBD-NL** | ***D*** **(Å) MBD-NL** |
|---|---|---|---|---|---|---|---|---|
| C-H | 0.68(10) | 2.72(5) | 0.99 | 2.69 | 0.99 | 2.76 | 1.00 | 2.74 |
| C-C | 0.83(10) | 2.64(5) | 0.99 | 2.64 | 0.98 | 2.68 | 0.98 | 2.68 |
| C-N | 0.69(10) | 2.53(5) | 0.94 | 2.56 | 0.90 | 2.58 | 0.87 | 2.60 |
| N | 0.76(10) | 2.38(5) | 0.81 | 2.47 | 0.74 | 2.44 | 0.66 | 2.46 |
| K | 0.76(10) | 3.75(10) | 1.00 | 3.50 | 0.94 | 3.19 | 1.00 | 3.40 |

**Table S4** Comparison of the experiential NIXSW structural parameters for the Ag(100) $K(TCNQ)_2$ phase with the predictions of the three different dispersion schemes.

| | ***f*** **expt** | ***D*** **(Å) expt** | ***f*** **vdW$^{surf}$-K$^{+}$** | ***D*** **(Å) vdW$^{surf}$-K$^{+}$** | ***f*** **vdW$^{surf}$** | ***D*** **(Å) vdW$^{surf}$** | ***f*** **MBD-NL** | ***D*** **(Å) MBD-NL** |
|---|---|---|---|---|---|---|---|---|
| C-H | 0.70(10) | 2.81(5) | 1.00 | 2.68 | 1.00 | 2.68 | 1.00 | 2.71 |
| C-C | 0.70(10) | 2.66(5) | 1.00 | 2.64 | 0.99 | 2.65 | 0.99 | 2.67 |
| C-N | 0.53(10) | 2.69(5) | 0.87 | 2.63 | 0.91 | 2.60 | 0.83 | 2.66 |
| N | 0.32(10) | 2.57(5) | 0.58 | 2.59 | 0.71 | 2.54 | 0.45 | 2.64 |
| K | 0.76(10) | 3.61(5) | 1.00 | 3.71 | 1.00 | 3.14 | 1.00 | 3.54 |

**Table S5** Comparison of the predicted NIXSW structural parameters for the Ag(100) KTCNQ phase obtained using the three different dispersion schemes.

| | ***f*** **expt** | ***D*** **(Å) expt** | ***f*** **vdW$^{surf}$-K$^{+}$** | ***D*** **(Å) vdW$^{surf}$-K$^{+}$** | ***f*** **vdW$^{surf}$** | ***D*** **(Å) vdW$^{surf}$** | ***f*** **MBD-NL** | ***D*** **(Å) MBD-NL** |
|---|---|---|---|---|---|---|---|---|
| C-H | - | - | 1.00 | 2.78 | 0.98 | 2.81 | 0.92 | 2.84 |
| C-C | - | - | 1.00 | 2.79 | 1.00 | 2.82 | 0.96 | 2.88 |
| C-N | - | - | 0.98 | 2.77 | 0.70 | 2.78 | 0.52 | 2.92 |
| N | - | - | 0.91 | 2.74 | 0.40 | 2.41 | 0.26 | 2.20 |
| K | - | - | 1.00 | 3.75 | 0.94 | 2.91 | 0.96 | 3.64 |

# 3. Theoretical Work function data

**Table S6** Work functions computed for both adsorbate systems and the clean surface. Reported are the work function $\phi$, the respective change in work function when compared to the clean substrate $\Delta\phi$, the electrostatic contribution of the work function change from the adsorbate overlayer $\Delta\mathrm{E}_{\mathrm{mol}}$, and the contribution of the work function change due to the chemical interaction of the overlayer with the metal $\Delta\mathrm{E}_{\mathrm{bond}}$. The two contributions sum up to the total work function change: $\Delta\phi = \Delta\mathrm{E}_{\mathrm{mol}} + \Delta\mathrm{E}_{\mathrm{bond}}$. $\Delta\mathrm{E}_{\mathrm{h}}$ and $\Delta E_e$ represent the hole and electron injection barriers, respectively. All values calculated at the PBE+vdW$^{surf}$(Cs/K$^+$)[2] level.

| | **Work function $\phi$ / eV** | **$\Delta\phi$ / eV** | **$\Delta E_{mol}$ / eV** | **$\Delta E_{bond}$ / eV** | **$\Delta E_h$ / eV** | **$\Delta E_e$ / eV** |
|---|---|---|---|---|---|---|
| Ag(100) | 4.19 | | | | | |
| $Cs(TCNQ)_4$ | 4.34 | 0.15 | -0.68 | 0.84 | 0.63 | 2.13 |
| $K(TCNQ)_4$ | 4.52 | 0.33 | -0.50 | 0.83 | 0.67 | 2.10 |
| $K(TCNQ)_2$ | 4.14 | -0.05 | -0.65 | 0.60 | 0.72 | 2.00 |
| KTCNQ | 3.28 | -0.91 | -1.27 | 0.36 | 0.80 | 1.86 |

**Table S7** Work functions computed for both adsorbate systems and the clean surface. Reported are the work function $\phi$, the respective change in work function when compared to the clean substrate $\Delta\phi$, the electrostatic contribution of the work function change from the adsorbate overlayer $\Delta E_{mol}$, and the contribution of the work function change due to the chemical interaction of the overlayer with the metal $\Delta E_{bond}$. The two contributions sum up to the total work function change: $\Delta\phi$ = $\Delta E_{mol}$+ $\Delta E_{bond}$. $\Delta E_h$ and $\Delta E_e$ represent the hole and electron injection barriers, respectively. All values calculated at the PBE+MBD-NL[4] level.

| | **Work function $\phi$ / eV** | **$\Delta\phi$ / eV** | **$\Delta E_{mol}$ / eV** | **$\Delta E_{bond}$ / eV** | **$\Delta E_h$ / eV** | **$\Delta E_e$ / eV** |
|---|---|---|---|---|---|---|
| Ag(100) | 4.19 | | | | | |
| $Cs(TCNQ)_4$ | 4.50 | 0.31 | -0.59 | 0.90 | - | - |
| $K(TCNQ)_4$ | 4.62 | 0.43 | -0.46 | 0.89 | - | - |
| $K(TCNQ)_2$ | 4.17 | -0.02 | -0.64 | 0.62 | - | - |
| KTCNQ | 3.94 | -0.25 | -0.71 | 0.47 | - | - |

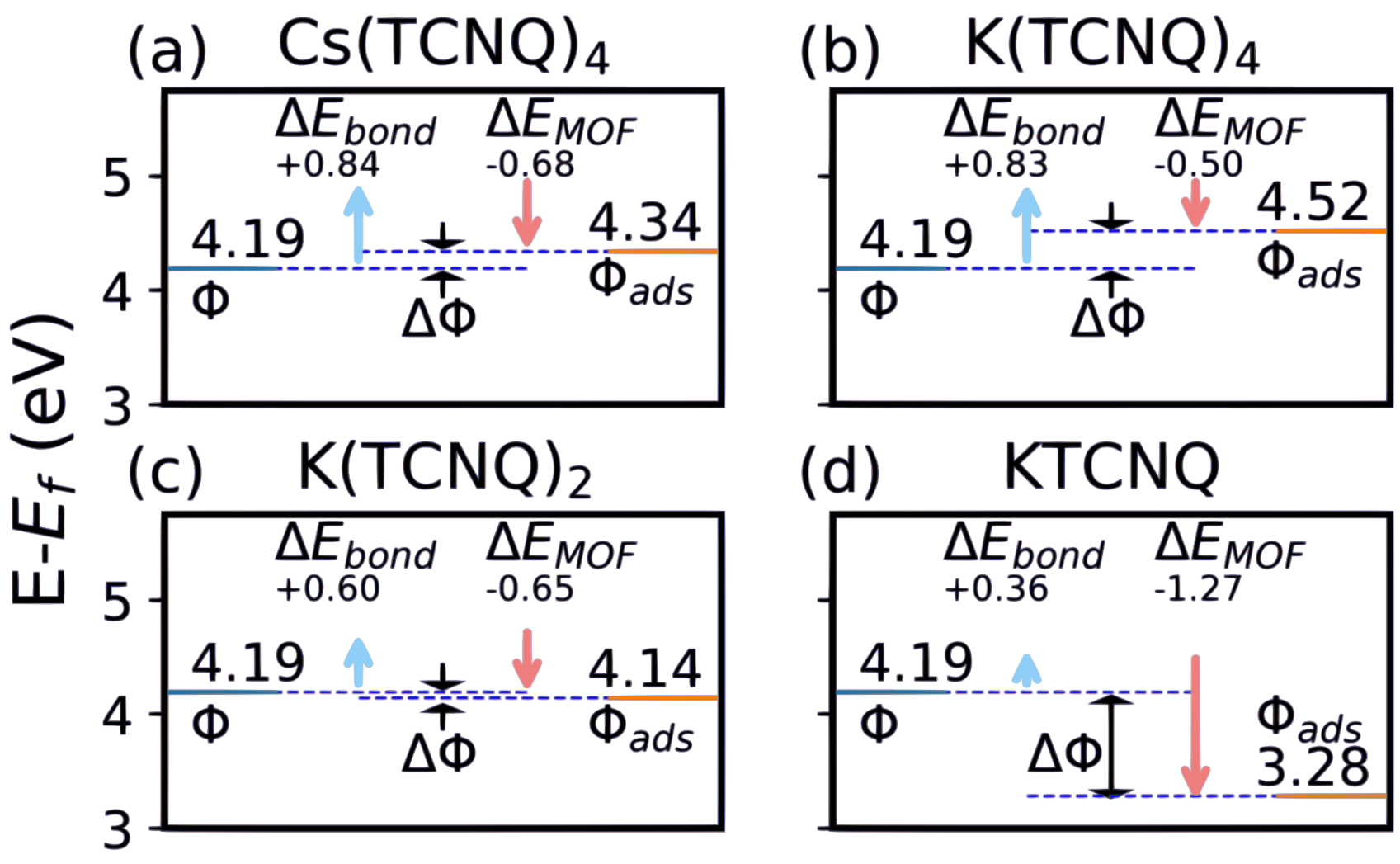

**Figure S4**- Schematic representation of the theoretical work function analysis. The work function of the clean surface, $\phi$, $\Delta E_{\mathrm{bond}}$(light blue), $\Delta E_{\mathrm{mol}}$ (light red), the work function after adsorption, $\phi_{\mathrm{ads}}$ and the change in work function $\Delta\Phi$. All values calculated at the PBE+vdW$^{\mathrm{surf}}$(Cs/$K^+$) level

## 4. Charge re-distribution data

**Table S8** - Hirshfeld charge[5] values for all species in the $Cs(TCNQ)_4$ unit cell. Units in elementary electronic charge *e*

| **$Cs(TCNQ)_4$** | **Cs (per atom)** | **TCNQ (per molecule)** | **Substrate (per unit cell)** |
|---|---|---|---|
| Hirshfeld charge | 0.51 | -0.80 | -2.68 |

**Table S9** - Hirshfeld charge[5] values for all species in the $K(TCNQ)_4$ unit cell. Units in elementary electronic charge *e*

| **$K(TCNQ)_4$** | **K (per atom)** | **TCNQ (per molecule)** | **Substrate (per unit cell)** |
|---|---|---|---|
| Hirshfeld charge | 0.36 | -0.75 | -2.64 |

**Table S10** - Hirshfeld charge[5] values for all species in the $K(TCNQ)_2$ unit cell. Units in elementary electronic charge *e*

| **$K(TCNQ)_2$** | **K (per atom)** | **TCNQ (per molecule)** | **Substrate (per unit cell)** |
|---|---|---|---|
| Hirshfeld charge | 0.40 | -0.91 | -1.42 |

**Table S11** - Hirshfeld charge[5] values for all species in the KTCNQ unit cell. Units in elementary electronic charge e

| **KTCNQ** | **K (per atom)** | **TCNQ (per molecule)** | **Substrate (per unit cell)** |
|---|---|---|---|
| Hirshfeld charge | 0.31 | -0.32 | -0.01 |

## 5. Molecular orbital density of states (MODOS)

Shown in Figure S5 are the MODOS for all systems calculated with PBE+vdW$^{surf}$(Cs/K$^+$). In the case of Cs(TCNQ)$_4$, the former LUMO+1, exhibits broader peaks than K(TCNQ)$_4$ which indicates a more complex hybridisation between the Ag and TCNQ states. All occupied states are very similar in both energetic position and shape. Furthermore, the larger absolute height of Cs compared to K can be partly attributed to the difference in van der Waals radii of each species but also to the increased dipole with increasing van der Waals radii. In the cases of K(TCNQ)$_2$ and KTCNQ, we find very little difference between the MODOS although a slight reduction in (former) HOMO and LUMO energetic positions.

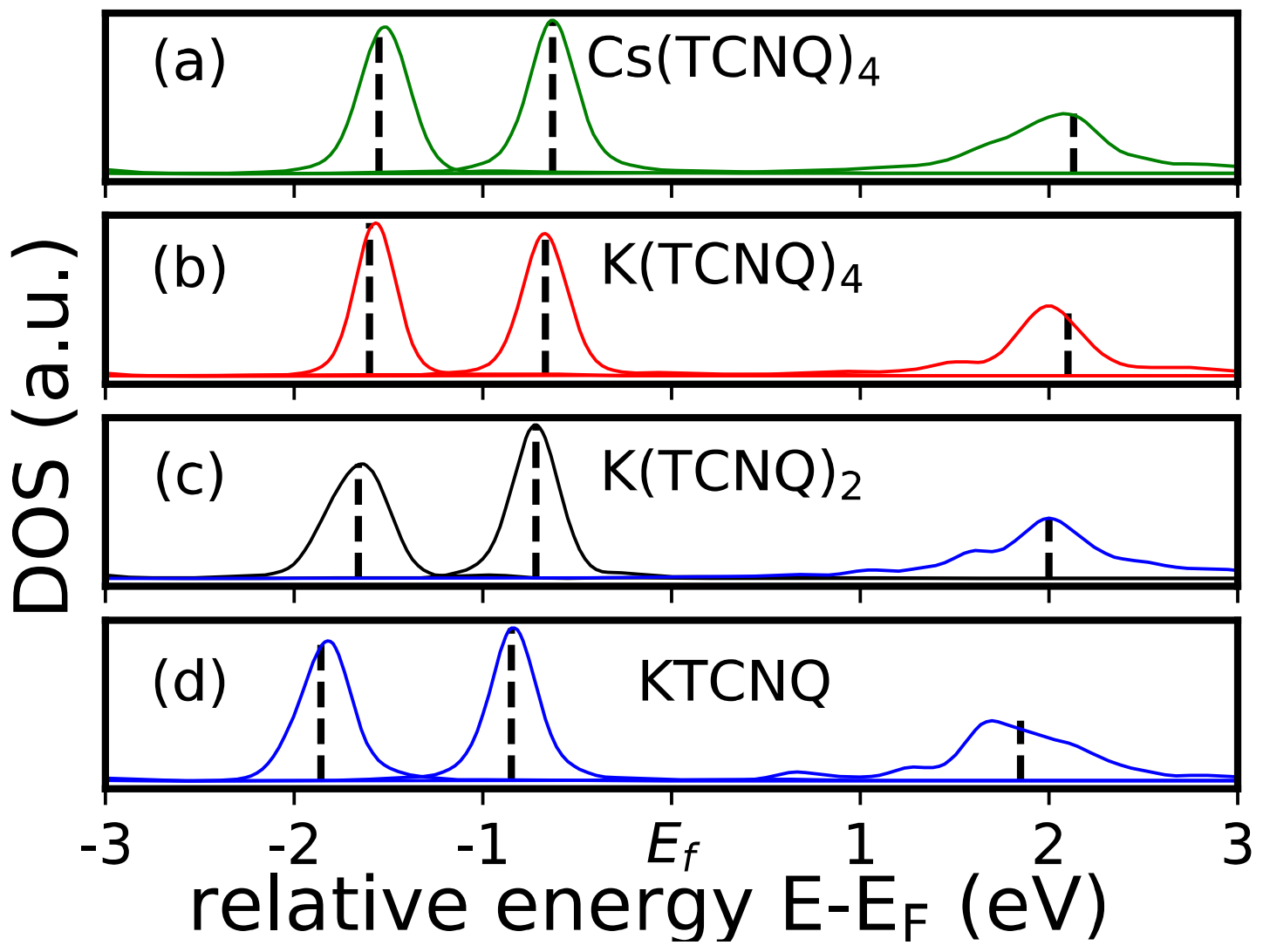

**Figure S5** - DOS plots of frontier states of TCNQ in the co-adsorbed system on Ag(100): (a) CsTCNQ$_4$, (b) KTCNQ$_4$, (c) KTCNQ$_2$, (d) KTCNQ (Left to right) filled peaks represent HOMO, and empty peaks represent the (former) LUMO and LUMO+1. Dashed black lines represent the energetic peak position as determined by equation 4 as reported in the main manuscript.

**References:**

(1) Hermann, K.; Van Hove, M. LEEDpat, 2014.

(2) Blowey, P. J.; Sohail, B.; Rochford, L. A.; Lafosse, T.; Duncan, D. A.; Ryan, P. T. P.; Warr, D. A.; Lee, T.-L.; Costantini, G.; Maurer, R. J.; Woodruff, D. P. Alkali Doping Leads to Charge-Transfer Salt Formation in a Two-Dimensional Metal–Organic Framework. *ACS Nano* **2020**, *14* (6), 7475–7483. https://doi.org/10.1021/acsnano.0c03133.

(3) Ruiz, V. G.; Liu, W.; Zojer, E.; Scheffler, M.; Tkatchenko, A. Density-Functional Theory with Screened van Der Waals Interactions for the Modeling of Hybrid Inorganic-Organic Systems. *Phys. Rev. Lett.* **2012**, *108* (14), 146103. https://doi.org/10.1103/PhysRevLett.108.146103.

(4) Hermann, J.; Tkatchenko, A. Density Functional Model for van Der Waals Interactions: Unifying Many-Body Atomic Approaches with Nonlocal Functionals. *Phys. Rev. Lett.* **2020**, *124* (14), 146401. https://doi.org/10.1103/PhysRevLett.124.146401.

(5) Hirshfeld, F. L. Bonded-Atom Fragments for Describing Molecular Charge Densities. *Theoret. Chim. Acta* **1977**, *44* (2), 129–138. https://doi.org/10.1007/BF00549096.